\documentclass[sigconf]{aamas} 

\usepackage{colortbl}  %彩色表格需要加载的宏包
\usepackage{xcolor}
\usepackage{listings}

\usepackage{threeparttable}
\usepackage{balance} % for balancing columns on the final page
\usepackage{amsmath}
\usepackage{cleveref}
\usepackage{enumitem}
\usepackage{tcolorbox}   % 彩色文本框 - 用于创建带边框的彩色任务框
\usepackage{xcolor}      % 颜色支持 - 用于定义和显示颜色
\usepackage{graphicx}    % 图片插入 - 用于插入图片
\usepackage{wrapfig}     % 浮动图形 - 用于图形环绕
\usepackage{caption}  
\usepackage{url}
\usepackage{hyperref}
\usepackage{ragged2e} 
\usepackage{multirow}
\usepackage{pifont}
\usepackage{subfig}
\usepackage{CJKutf8}
\usepackage{balance}   % 在导言区引入宏包
\usepackage{tcolorbox}
\usepackage{subcaption}

\doi{HJFB4234}

\makeatletter
\gdef\@copyrightpermission{
  \begin{minipage}{0.2\columnwidth}
   \href{https://creativecommons.org/licenses/by/4.0/}{\includegraphics[width=0.90\textwidth]{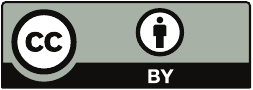}}
  \end{minipage}\hfill
  \begin{minipage}{0.8\columnwidth}
   \href{https://creativecommons.org/licenses/by/4.0/}{This work is licensed under a Creative Commons Attribution International 4.0 License.}
  \end{minipage}
  \vspace{5pt}
}
\makeatother

\setcopyright{ifaamas}
\acmConference[AAMAS '26]{Proc.\@ of the 25th International Conference
on Autonomous Agents and Multiagent Systems (AAMAS 2026)}{May 25 -- 29, 2026}
{Paphos, Cyprus}{C.~Amato, L.~Dennis, V.~Mascardi, J.~Thangarajah (eds.)}
\copyrightyear{2026}
\acmYear{2026}
\acmDOI{}
\acmPrice{}
\acmISBN{}

\title[Automatically Benchmarking LLM Code Agents through Agent-Driven Annotation and Evaluation]{Automatically Benchmarking LLM Code Agents through Agent-driven Annotation and Evaluation}

\author{Lingyue Fu}
\authornote{Work done while intern at Meituan.}
\affiliation{
  \institution{Shanghai Jiao Tong University}
  \city{Shanghai}
  \country{China}}

\author{Bolun Zhang}
\authornotemark[1]
\affiliation{
  \institution{Shanghai Jiao Tong University}
  \city{Shanghai}
  \country{China}}

\author{Hao Guan}
\authornotemark[1]
\affiliation{
  \institution{Shanghai Jiao Tong University}
  \city{Shanghai}
  \country{China}}

  \author{Yaoming Zhu}
\affiliation{
  \institution{Meituan}
  \city{Shanghai}
  \country{China}}

  \author{Lin Qiu}
\affiliation{
  \institution{Meituan}
  \city{Shanghai}
  \country{China}}

\author{Weiwen Liu}
\authornote{Corresponding Author. wwliu@sjtu.edu.cn, wnzhang@sjtu.edu.cn, yyu@sjtu.edu.cn.}
\affiliation{
  \institution{Shanghai Jiao Tong University}
  \city{Shanghai}
  \country{China}}
  
    \author{Xuezhi Cao}
\affiliation{
  \institution{Meituan}
  \city{Shanghai}
  \country{China}}

    \author{Xunliang Cai}
\affiliation{
  \institution{Meituan}
  \city{Shanghai}
  \country{China}}
  
\author{Weinan Zhang}
\authornotemark[2]
\affiliation{
  \institution{Shanghai Jiao Tong University}
  \city{Shanghai}
  \country{China}}

  \author{Yong Yu}
  \authornotemark[2]
\affiliation{
  \institution{Shanghai Jiao Tong University}
  \city{Shanghai}
  \country{China}}

\begin{abstract}
Recent advances in code agents have enabled automated software development at the project level, supported by large language models (LLMs). However, existing benchmarks for code agent evaluation face two major limitations. First, creating high-quality project-level evaluation datasets requires extensive domain expertise, leading to prohibitive annotation costs and limited diversity. Second, while recent Agent-as-a-Judge paradigms address the rigidity of traditional unit tests by enabling flexible metrics, their reliance on In-Context Learning (ICL) with general LLMs often results in inaccurate assessments that misalign with human standards. To address these challenges, we propose an agent-driven benchmark construction pipeline that leverages human supervision to efficiently generate diverse project-level tasks. Based on this, we introduce PRDBench, comprising 50 real-world Python projects across 20 domains, each with structured Product Requirement Documents (PRDs) and comprehensive criteria. Furthermore, to overcome the inaccuracy of general LLM judges, we propose a highly reliable evaluation framework powered by a specialized, fine-tuned model. Based on Qwen3-Coder-30B, our dedicated PRDJudge achieves over 90\% human alignment in fixed-interface scenarios. Extensive experiments demonstrate that our suite provides a scalable, robust, and highly accurate framework for assessing state-of-the-art code agents.
\end{abstract}


\newcommand{\BibTeX}{\rm B\kern-.05em{\sc i\kern-.025em b}\kern-.08em\TeX}

\begin{document}

%%% The following commands remove the headers in your paper. For final 
%%% papers, these will be inserted during the pagination process.

\pagestyle{fancy}
\fancyhead{}

%%% The next command prints the information defined in the preamble.

\maketitle 

%%%%%%%%%%%%%%%%%%%%%%%%%%%%%%%%%%%%%%%%%%%%%%%%%%%%%%%%%%%%%%%%%%%%%%%%

\section{Introduction}
In recent years, code agents have made significant progress, capable of solving increasingly complex programming tasks. From initially focusing on single-file code generation to now advancing towards complete project-level software development, Code Agents are demonstrating powerful capabilities in building comprehensive systems. LLM-based Code Agents such as CursorAgent~\cite{anysphere2025cursoragent}, Claude Code~\cite{anthropic2025claude37} and Gemini CLI~\cite{mullen2025geminicli} have been widely adopted in real-world development scenarios, accelerating the proliferation of automated programming tools.

As code agents continue to advance, a variety of benchmarks have emerged to evaluate their capabilities. Several works~\cite{chan2024mlebench,huang2024dacodeagentdatascience,daghighfarsoodeh2025deepbenchdeeplearningbenchmark} focus on assessing LLMs and code agents in data science tasks, including subtasks such as data processing and machine learning training. Other benchmarks~\cite{jimenez2023swe,fu2025corecodebench,vergopoulos2025automatedbenchmarkgenerationrepositorylevel} select complex code repositories from GitHub pull requests and transform them, often with minimal human intervention, into testable scenarios. In addition, PaperBench~\cite{starace2025paperbench} evaluates the ability of agents to reproduce paper-level code implementations, providing a more rigorous assessment of their research reproducibility.

\begin{table*}[t]
    \centering
    \begin{tabular}{lcccccc}
        \toprule
        \textbf{Benchmark} & \textbf{Area} & \textbf{\# Project} & \textbf{Annotator} & \textbf{\# Metrics} & \textbf{Judger}  & \textbf{Claude Code Score}\\ \midrule
         SWE-Bench~\cite{jimenez2023swe} & Pull Requests & 12 & Web Crawler & 2,294 & Unit Test  & 70.3\%$^*$\\
         MLEBench~\cite{chan2024mlebench} & AI Competition & 75 & Human & 75 & Test Set & 51.1\% \\
        DevAI~\cite{zhuge2024agent} &  Developement & 55 & Human & 365  & Agent  & 73.0\%$^\dagger$ \\
        PaperBench~\cite{starace2025paperbench} & AI Research & 20 & Human (PhD.) & 8,316 & Human/LLM &21.0\%  \\ \midrule
        PRDBench & Engineering Development & 50 & Agent \& Human & 1,258 &  Agent (PRDJudge) &  45.5\%\\ \bottomrule
    \end{tabular}
    \caption{Comparison between PRDBench and project-level code agent benchmarks. We employ the performance of code agents with Claude as a benchmark indicator for assessing the task difficulty of our proposed evaluation suite. The results from external report~\cite{anthropic2025claude37} is denoted by $*$, from our implementation is denoted by $\dagger$, otherwise we use the best results from the original paper with author specified Claude-driven agents.}
    \label{tab:comparison}
     % \vspace{-20pt}
\end{table*}

Despite these advancements, current benchmarks face two major limitations. First, \textit{the creation of high-quality evaluation datasets requires substantial human effort and expertise}. Generating reliable unit tests or grading tasks often depends on expert annotators, especially when migrating real-world scenarios such as Kaggle competitions or GitHub pull requests. These tasks demand domain-level verification and annotation to produce usable test cases. As code agents advance rapidly, the requirements for annotators' expertise and time investment continue to increase. For example, PaperBench~\cite{starace2025paperbench} recruited ICML authors as annotators, with each annotation task demanding several days of work from PhD-level experts. This process not only incurs high annotation costs, but also restricts the diversity of evaluation data, as recruiting experts from various domains is challenging and often leads to datasets being sourced from a single domain.

Second, \textit{the rigidity of the evaluation limits the broad applicability of current benchmarks}. Existing benchmarks~\cite{jimenez2023swe} primarily rely on unit test pass rates to assess code agent performance. While unit tests are effective for verifying specific functions or components, they are limited in scope and overly restrictive, enforcing strict requirements on project interfaces and implementation details. To address this rigidity and introduce broader validation methods, recent works have explored LLM-as-a-Judge or Agent-as-a-Judge paradigms. However, these approaches predominantly rely on In-Context Learning (ICL), using zero-shot or few-shot prompts to guide general-purpose LLMs for specific scoring tasks. Relying solely on ICL for complex, project-level evaluation often leads to inaccurate assessments. General LLMs struggle to correctly interpret intricate execution logs or subtle file differences, resulting in significant misalignment with human judgments. For instance, the LLM judger introduced in PaperBench~\cite{starace2025paperbench} achieves a maximum human alignment rate of only 83\%. Thus, achieving both evaluation flexibility and high scoring accuracy remains a critical challenge.

To address the two aforementioned challenges, we propose an agent-driven construction approach that enables the creation of project-level benchmarks with flexible metrics at low human cost. Specifically, a state-of-the-art code agent is used to generate both the project scaffolding and a Product Requirement Document (PRD) along with an executable criteria scheme. Human annotators are only required to verify whether the criteria scheme is compatible with the scaffolding interfaces and whether the expected outputs are reasonable, without the need to manually create detailed evaluation standards or reference solutions. This significantly reduces annotation complexity: for PRDBench, annotators with undergraduate-level knowledge in software engineering related fields are able to complete the annotation, with an average of only eight hours needed to finish the scaffolding and metrics for each project, greatly improving annotation efficiency.

Based on this approach, we build a Product Requirement Document~(PRD)-centered benchmark, named \textbf{PRDBench}\footnote{Source of PRDBench is available in \url{https://github.com/AGI-Eval-Official/PRDBench}.}. PRDBench consists of 50 real-world coding tasks, each defined by a structured PRD, a well-specified and verifiable criteria scheme, and a standard solution code repository. The coding tasks are sourced from real-world requirements, academic projects, and thesis work, spanning 20 common subdomains. For each task, the PRD provides a criteria scheme that facilitates comprehensive human-like quality assurance~(QA) checks. We present a comparison between PRDBench and previous code agent benchmarks in Table~\ref{tab:comparison}, where PRDBench offers a comprehensive, multifaceted, project-level benchmark for code agents. 
During evaluation, we employ the Agent-as-a-Judge paradigm to score code according to the criteria scheme, enabling the evaluation of a wide range of test types beyond unit tests. However, relying on off-the-shelf general LLMs as judges often leads to instability and hallucinations in complex project-level assessments. To ensure highly reliable evaluation, we further introduce a specialized, fine-tuned PRDJudge based on Qwen3-Coder-30B. By training on a curated dataset of evaluation trajectories, our dedicated PRDJudge overcomes the limitations of general models, achieving an over 90\% human alignment rate in fixed-interface scenarios.

In summary, our contributions include:
\begin{itemize}[leftmargin=10pt]
    \item We design an \textbf{agent-driven data production pipeline}, where human supervision guides agents to efficiently generate challenging test cases that go beyond current agents capabilities. This pipeline alleviates the need for expertise annotators.
    \item We construct \textbf{PRDBench}, which covers 20 common domains in Python software development. Each benchmark item includes a PRD and evaluation metrics, and we design three categories of agent-friendly test types for comprehensive evaluation.
    \item We adopt the \textbf{Agent-as-a-Judge paradigm} powered by a specialized fine-tuned model. Based on Qwen3-Coder-30B, our \textbf{PRDJudge}\footnote{Model weights are publicly available at: \url{https://huggingface.co/AGI-Eval/PRDjudge}} achieves over \textbf{90\%} human alignment in fixed-interface scenarios, enabling flexible, human-like QA assessment beyond traditional unit tests.
    \item We conduct extensive experiments on PRDBench, detailing the training process and data construction of the evaluation agent, and providing useful insights into both state-of-the-art code agents and the evaluation paradigm.
\end{itemize}
% The most recent version including full appendix and additional results is maintained at \url{https://arxiv.org/abs/2510.24358}.

\begin{figure*}[t]
    \centering
    \includegraphics[width=0.95\linewidth]{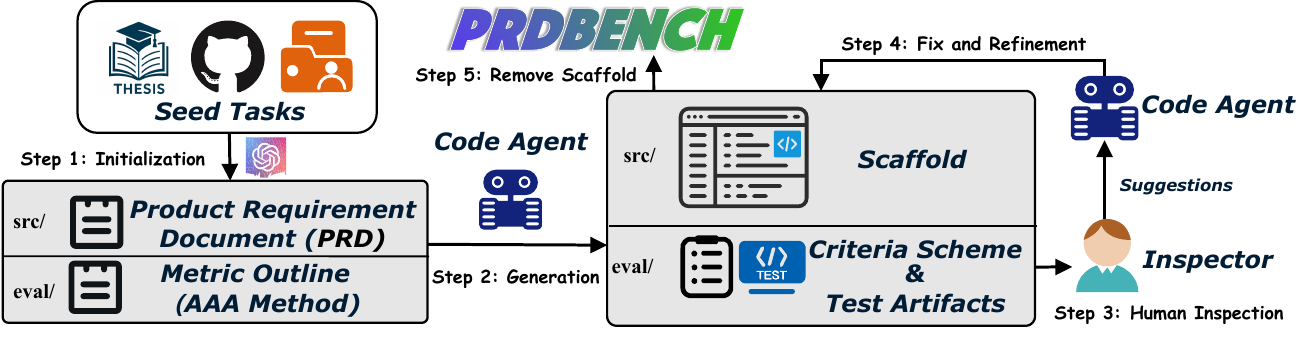}
    \caption{\textbf{Overview of the PRDBench data production workflow.} Step1: PRD and Test Plan Initialization; Step 2: Code Scaffold and Criteria Scheme Generation; Step 3: Human Inspection; Step 4: Agent-based Fix and Refinement; Step 5: Remove scaffold.}
    \label{fig:generate}
    % \vspace{-5pt}
\end{figure*}

\section{Related Work}
\subsection{Code Agent Evaluation}
% project eval :user interaction simulation for execution
% xiang2025scireplicatebenchbenchmarkingllmsagentdriven: execution accuracy, CodeBLEU, and repository dependency/API recall metrics.

To keep pace with the rapid development of LLMs' and agents' coding capabilities, recent benchmarks increasingly emphasize the construction of executable and complex tasks within real software projects. These benchmarks are typically built by mining human-generated data from online platforms or through extensive manual annotation. Some benchmarks~\cite{li2024promptinglargelanguagemodels, huang2024dacodeagentdatascience} require annotators to manually create test points for each step, while others~\cite{jimenez2024swebenchlanguagemodelsresolve, yang2024execrepobenchmultilevelexecutablecode, pan2024codevbenchllmsunderstanddevelopercentric, wu2024repomasterevalevaluatingcodecompletion, hai2025impactscontextsrepositorylevelcode} rely on authentic submission records, using the inherent difficulty of submitted code as a basis for evaluation. Additionally, some benchmarks~\cite{li2024evocodebenchevolvingcodegeneration,zhuo2025bigcodebenchbenchmarkingcodegeneration,peng2024humanevalxlmultilingualcodegeneration} leverage existing resources and require expert-driven rewriting, which demands significant manual effort. 
As code agents become more capable, these annotation paradigms require increasingly specialized expertise, making the recruitment of expert annotators for each domain costly and limiting the scalability and diversity of benchmark construction.

In terms of evaluation metrics, most existing benchmarks adopt relatively narrow criteria, such as unit test pass rates~\cite{jimenez2023swe, li2024promptinglargelanguagemodels, xiang2025scireplicatebenchbenchmarkingllmsagentdriven}, which primarily assess functional correctness. ProjectEval~\cite{liu2025projectevalbenchmarkprogrammingagents} simulates user interaction to evaluate projects, focusing on specific types of tasks. However, these evaluation criteria are limited to specific task types and cannot comprehensively assess diverse, complex software projects.

\begin{figure*}[t]
\centering
    \includegraphics[width=0.9\linewidth]{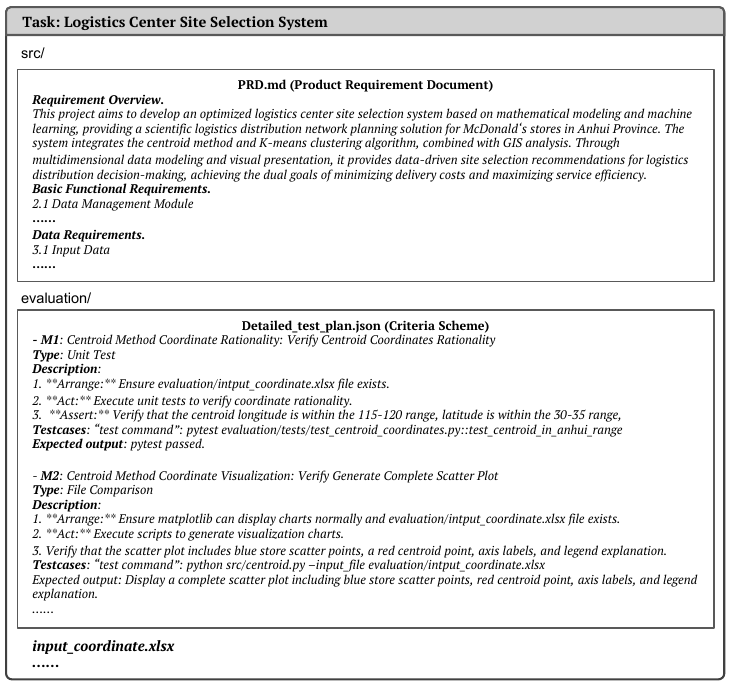}
    \caption{A task example of PRDBench.}
    \label{fig:demo}
\end{figure*}

\subsection{Agent-as-a-Judge}
% wang2025codevisionaryagentbasedframeworkevaluating   LLM-based agent framework for evaluating LLMs in code generation
% 

Due to the inherent limitations of unit test-based benchmarks, recent studies have begun to explore the use of LLMs as judges for agent evaluation, a paradigm often referred to as LLM-as-a-Judge. LLMs can provide capabilities such as image recognition~\cite{chen2024mllm}, code alignment~\cite{weyssow2024codeultrafeedbackllmasajudgedatasetaligning}, and question-answering~\cite{ho2025llm}, breaking the restriction that benchmark metrics must be executable. With the emergence of agents equipped with tools, LLMs are able to read large volumes of files and perform extensive analyses, offering new possibilities for evaluation. DevAI~\cite{zhuge2024agent} is the first work to apply the agent-as-a-judge paradigm to code agent evaluation. It focuses on checking program execution and output format, but lacks comprehensive assessment of overall engineering effectiveness. Some studies\cite{wang2025codevisionaryagentbasedframeworkevaluating}, have explored LLM-based agent frameworks for evaluating code generation, improving the accuracy of Agent-as-a-Judge through architectural innovations. However, the lack of well-defined metrics and code agents has limited the use of agent-as-a-judge. PRDBench fills this gap with structured criteria and dedicated tools, enabling more robust evaluation.

\subsection{Fine-tuned LLMs as Judges}
Recent studies suggest that for most tasks, evaluating outputs is inherently easier than generating them~\cite{together2024finetuning}, motivating a line of work that fine-tunes smaller open-source LLMs as dedicated judges. PandaLM~\cite{wang2024pandalmautomaticevaluationbenchmark}, Prometheus~\cite{kim2024prometheusinducingfinegrainedevaluation}, JudgeLM~\cite{zhu2025judgelmfinetunedlargelanguage}, and Auto-J~\cite{li2023generativejudgeevaluatingalignment} demonstrate that fine-tuned models of 7B–33B parameters can match or even surpass GPT-4 on targeted evaluation tasks, while offering significant advantages in cost, speed, and privacy. Notably, Huang et al.~\cite{huang2025empiricalstudyllmasajudgellm} further reveal that such fine-tuned judges essentially operate as highly effective task-specific classifiers, consistently achieving state-of-the-art performance on in-domain benchmarks. These findings motivate our approach of fine-tuning a dedicated PRDJudge specifically for project-level code evaluation, where domain-specific expertise is critical for accurate and reliable scoring.

\section{PRDBench}
% In this section, we introduce PRDBench, a comprehensive benchmark designed to evaluate the end-to-end development capabilities of code agents based on real-world PRDs. We detail the benchmark's construction process and present a statistical overview of the final dataset.

\subsection{Seed Tasks}
Our seed tasks are carefully curated from three primary sources to ensure diversity and real-world relevance: (1) real-world user development requests collected from an internal AI product development platform; (2) publicly available final projects from Computer Science (CS) courses hosted on GitHub; and (3) code reproduction tasks derived from academic theses in the CS domain. To ensure the suitability and consistency of tasks for PRDBench, we apply a rigorous filtering process. Specifically, we require that each selected task can be fully implemented in Python and that all its associated datasets are publicly accessible. We choose Python as the programming language for PRDBench primarily due to its versatility and comprehensive ecosystem, which supports a wide range of programming paradigms and application domains. This ensures that PRDBench tasks are representative of diverse, real-world scenarios.

\subsection{Agent-Driven Data Production}
Figure~\ref{fig:generate} illustrates the agent-driven data production workflow of PRDBench. Throughout the annotation process, we apply SOTA code agents to generate the project scaffold and criteria scheme.  Human annotators are only required to supervise the quality of criteria scheme, i.e., whether the criteria aligns with the scaffold interfaces and whether the expected outputs meet the requirements specified in the PRD. 
It is worth mentioning that manual metric annotation and code modification are not necessary, which greatly reduces both the complexity and time required for annotation.

\textbf{Step~1:~PRD and Test Plan Initialization.} 
After selecting seed tasks, we utilize code agents (such as Claude Code) to generate detailed and standardized PRD documents, ensuring clarity and completeness in task specification. The PRD serves as the evaluation blueprint for PRDBench, and includes sections such as Requirement Overview, Functional Requirements, and Data Requirements. Based on the PRD, we employ GPT-4.1 to generate a corresponding metric outline, structuring test cases using the Arrange-Act-Assert (AAA) methodology~\cite{wei2025developers}. The Arrange step sets up the test case by preparing the necessary files, input data, and environment configurations. The Act step focuses on the core behavior to be tested, such as running the program or performing interaction tests to obtain output results. The Assert step verifies the expected outcomes by checking the system’s response or state, ultimately determining whether the test passes or fails. This structure is applicable to a wide range of code-related testing scenarios, providing PRDBench with a  executable testing plan.

\begin{figure}[t]
        \centering
        \includegraphics[width=\linewidth]{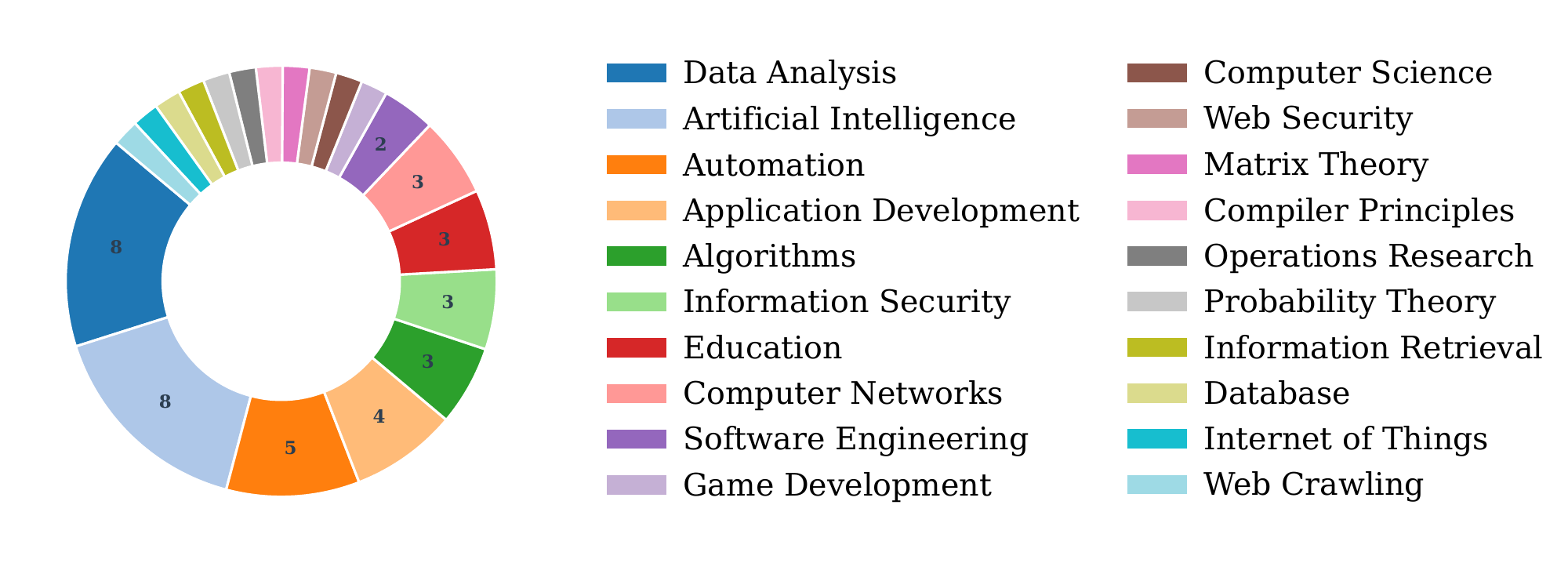}
        \caption{Domain distribution of PRDBench.}
        \label{fig:sub_a}
\end{figure}

\textbf{Step~2:~Scaffold and Criteria Generation.}
After initializing with the PRDs and metric outlines, we employ state-of-the-art code agents to generate the code scaffold for each task. The scaffold includes both module design and interface design, serving as the core framework of the entire project. In some cases, particularly for complex algorithms, the implementation may not be fully correct as long as it does not affect the annotation process. Building upon the scaffold, we further utilize code agents to expand and refine the metric outline into a specific criteria scheme, which includes necessary test interfaces and test artifacts. The presence of well-defined code interfaces in the scaffold simplifies and standardizes the generation of the criteria scheme, making the process more efficient and consistent.

\textbf{Step~3: Human Inspection.} 
With the scaffold and the criteria scheme in place, the inspection process for human annotators becomes straightforward. Annotators conduct inspection only by running the tests to verify whether the interfaces function correctly and whether the expected outputs in the criteria scheme align with the PRD requirements. Notably, even for tasks sourced from expertise projects or academic papers, annotators only need basic computer science knowledge to perform effective annotation. 

\textbf{Step~4:~Agent-Based Fix and Refinement.} If any issues are identified during step 3, annotators shall provide targeted feedback to the code agent for refining the scaffold or criteria scheme. The code agent then revises the relevant components, and the inspection process is repeated. This iterative process continues until all issues are resolved. To ensure the rigorous quality and sufficient complexity of our benchmark, we establish a strict inclusion criterion: only tasks that have undergone at least 5 rounds of such iterative human-agent refinement are retained.

\textbf{Step~5:~Remove Scaffold.}
 Finally, the scaffold is removed, retaining only the PRD, criteria scheme, test artifacts, and data. This forces the evaluated agents to generate code from scratch, enabling a rigorous assessment of their end-to-end development capabilities.

An overall example of a PRDBench task is illustrated in Figure~\ref{fig:demo}. Each PRDBench testcase consists of a PRD document and an evaluation suite, which includes a criteria scheme and test artifacts.

\subsection{Data Statistics}
PRDBench comprises 50 project-level Python tasks that span 20 distinct application domains. To ensure broad and balanced coverage, we utilized GPT-4 to assign domain labels to all candidate tasks before strategically sampling the final dataset. As illustrated in Figure~\ref{fig:sub_a}, we include at least one representative task from less common domains. For more prevalent areas, such as data processing and machine learning, we select tasks that demand a diverse array of technical approaches and sub-skills. This sampling strategy ensures that PRDBench accurately represents diverse, real-world Python development scenarios.

\begin{figure}[t]
        \centering
        \includegraphics[width=0.8\linewidth]{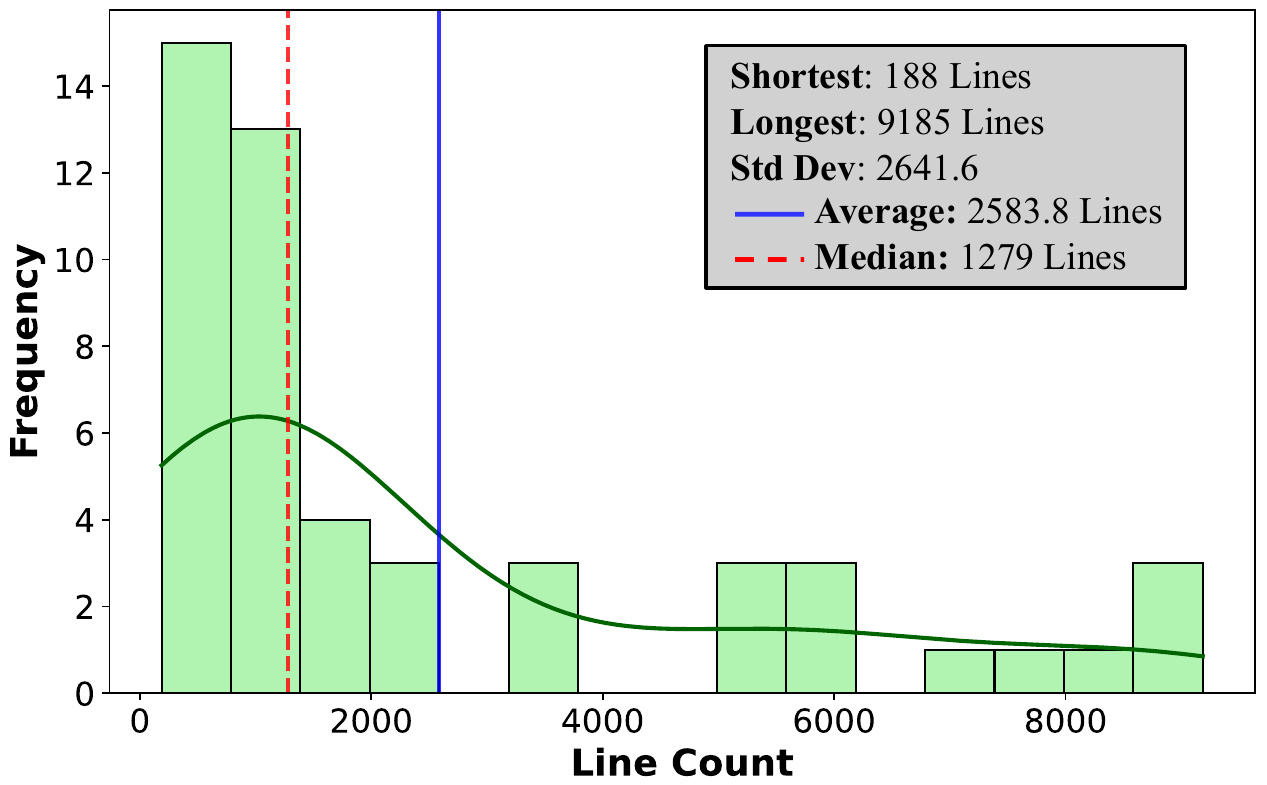} 
        \caption{Scaffold line counts distribution.} % <--- 必须有 caption，否则 label 无法正确编号
        \label{fig:sub_b}
\end{figure}

% 2. 任务规模：PRD与代码复杂度
These tasks exhibit substantial complexity. On average, each structured PRD contains 105.22 lines of detailed requirements. Furthermore, Figure~\ref{fig:sub_b} presents the distribution of scaffold line counts recorded during the annotation process, highlighting the realistic scale of the projects.

% 3. 评估维度：测试指标分类
For comprehensive evaluation, PRDBench encompasses a total of 1,258 evaluation metrics. These metrics collectively simulate the different stages of quality assurance (QA) in real-world software engineering and are categorized into three main types:
\begin{itemize}[leftmargin=10pt] 
\item \textbf{Unit Test (408 metrics):} As in previous benchmark studies~\cite{jimenez2023swe}, unit tests remain the most effective approach for verifying the functional correctness of individual components and modules. PRDBench retains this rigorous testing paradigm and provides auxiliary \texttt{pytest}-based test scripts. 
\item \textbf{Shell Interaction (732 metrics):} For tasks involving command-line interaction or external system operations, predefined shell commands are executed to compare actual outputs against expected results. This ensures the code correctly handles system-level operations and user inputs. PRDBench supplies comprehensive simulated user input files and program entry commands for this purpose. 
\item \textbf{File Comparison (118 metrics):} For project-level tasks that generate files or require specific directory structures, the produced artifacts are compared against reference solutions to verify correctness in content, format, and organization. PRDBench provides the raw data for target file generation, the corresponding Python commands, and the reference solution files. 
\end{itemize}

\section{PRDJudge}
In this section, we present PRDJudge, a specialized evaluation agent developed to automatically and reliably assess code submissions on PRDBench. We elaborate on its tool-augmented evaluation framework, and the fine-tuning methodology designed to align its scoring with human standards.

\subsection{Evaluation Framework}

\begin{figure}[t]
    \centering
    \includegraphics[width=0.9\linewidth]{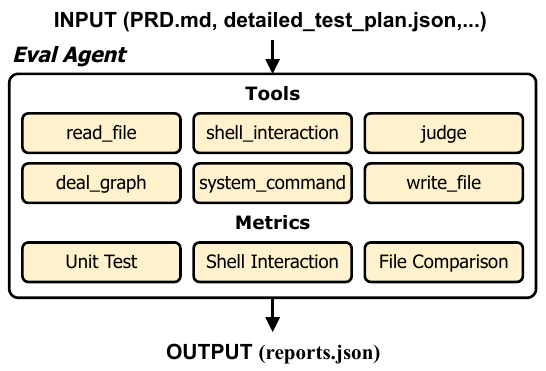}
    \caption{Overview of PRDJudge. PRDJudge executes tests based on the criteria scheme using various tools, compares outputs (files or results) with expected outputs, and generates a report for the submitted code.
}
    \label{fig:evalagent}
\end{figure}

To automate the validation of code generated by code agents, we introduce \textbf{PRDJudge}, a lightweight evaluation agent tailored for the predefined criteria schemes of PRDBench (illustrated in Figure~\ref{fig:evalagent}). For each item in the criteria scheme, PRDJudge leverages the executable terminal commands and expected outputs provided by the benchmark to conduct systematic verification. 

To interact effectively with the testing environment, PRDJudge is equipped with six core tools, including file reading and writing, command-line execution, image handling, and a specialized judge tool. Among these, the \texttt{deal graph} tool offers a multimodal LLM interface (powered by GPT-4o in this paper), allowing the agent to process visual outputs by submitting both the image and a descriptive prompt to obtain verification results—this is particularly useful for tasks requiring UI or plot validation. Furthermore, the judge tool is specifically designed for PRDBench. It accepts files containing simulated user inputs, enabling the generation of comprehensive terminal logs for thorough analysis without the need for manual input specification.

By leveraging this toolset, PRDJudge automatically adapts to specific code repositories and seamlessly executes the three categories of tests (Unit Test, Shell Interaction, and File Comparison) defined in Section~3.3. Since all supporting files and test categories are self-contained within PRDBench, the complexity of evaluation is greatly reduced. PRDJudge only needs to execute each test, analyze the execution logs or file differences, and generate a detailed report on the performance and correctness of the code agent submissions. However, accurately interpreting complex execution logs and aligning the final scores with human quality assurance standards requires exceptional reasoning capabilities, which motivates the training of a specialized evaluation model, as detailed in the following section.

\subsection{Fine-tuning PRDJudge}

\subsubsection{Human Annotation Protocol}
To establish a high-quality ground truth for both training data filtering and final evaluation, we employ a rigorous human annotation process. Detailed information of our annotators is provided in Appendix~\ref{sec:background}. Each scoring point in the benchmark is independently evaluated by two human annotators. In cases of disagreement, a third senior annotator serves as an arbitrator to make the final decision. The annotation follows a strict three-tier ordinal scoring system:
\begin{itemize}[leftmargin=10pt]
    \item \textbf{Score 2 (Pass):} The code executes perfectly, and the output strictly matches the expected requirements specified in the criteria scheme.
    \item \textbf{Score 1 (Partial):} The code runs successfully (i.e., the interfaces are correct and no execution crashes occur), but the final output or behavior does not fully align with the expected results.
    \item \textbf{Score 0 (Fail):} The code fails to execute entirely, typically due to syntax errors, missing dependencies, or severe runtime exceptions.
\end{itemize}

\subsubsection{Training Data Construction}
The training data for PRDJudge consists of high-quality evaluation trajectories. To collect these trajectories, we deploy a highly capable model, Qwen3-Coder-480B-A3B~\cite{qwen3technicalreport}, as the initial judge within the PRDJudge framework. We then apply this judge to score code submissions from 11 code agents spanning diverse backbone LLMs and agent frameworks, collecting the resulting evaluation trajectories as candidate training data. To ensure diversity in the training distribution, we randomly sample 8 repositories for each agent to form the training pool.

Crucially, to prevent the model from learning hallucinated or flawed reasoning, we apply a strict two-stage filtering mechanism. Initially, Qwen3-Coder-480B-A3B generated a total of 2,147 evaluation trajectories. In the first stage, we only retain the 1,742 trajectories where the final score predicted by the model exactly matches our human-annotated ground truth. It is worth noting that while the generation of PRDBench tasks is highly efficient and scalable (as discussed in Section 1), establishing this ground truth required a substantial human annotation effort. However, we consider this intensive labeling process a necessary, one-time investment to build a reliable automated judge, which subsequently enables zero-cost, scalable evaluations for any future code agents. Furthermore, to ensure the efficiency and accuracy of the agent's testing trajectories, we apply an additional rule-based filter to eliminate trajectories that are excessively long or exhibit invalid tool usage. Although these discarded trajectories achieved the correct final score, their inaccurate tool usage could introduce noise into the evaluation agent's scoring behavior. Ultimately, this rigorous filtering process yields a final dataset of 911 high-quality training trajectories for fine-tuning.

\section{Experiments and Results}
% In this section,

\subsection{Code Agents}

\begin{table}[t]
\centering
\small
\caption{Agent Specifications and Open-Source Status.}
\label{tab:agent_models}
\resizebox{\linewidth}{!}{%
\begin{tabular}{lllcc}
\toprule
\textbf{Agent Type} & \textbf{Agent Framework} & \textbf{Model} & \textbf{Agent OSS} & \textbf{LLM OSS} \\
\midrule
\multirow{8}{*}{\centering Minimal}  
    & \multirow{8}{*}{Basic code tools}    & Qwen3-Coder-480B-A3B~\citep{qwen3technicalreport}         &   \checkmark  & \checkmark \\
    &   & GPT-5.2~\cite{gpt5.2}    
    &  \checkmark   & \ding{55} \\
    &   & Claude-4.5-Sonnet~\cite{anthropicclaude45} 
    &  \checkmark   & \ding{55} \\
    &   & Gemini-3-pro~\cite{gemini3}        
    &  \checkmark   & \ding{55} \\
    & & 
Kimi-K2-Instruct-0905~\cite{kimiteam2025kimik2openagentic}  & \checkmark   & \checkmark \\
    & & DeepSeek-V3.2%~\citep{deepseekai2025deepseekv32} 
    & \checkmark   & \checkmark\\
    & & GLM-4.7~\cite{5team2025glm45agenticreasoningcoding} 
    & \checkmark   & \checkmark \\
    & & Minimax-M2~\cite{minimaxm2} 
    & \checkmark   & \checkmark\\
\midrule
\multirow{4}{*}{\centering Commercial}    
    & Gemini CLI~\cite{mullen2025geminicli} & Gemini-2.5-pro~\cite{gemini25}     & \checkmark    & \ding{55} \\
    & Claude Code~\cite{anthropic2025claude37} & Claude-4.5-Sonnet~\cite{anthropicclaude45}  & \ding{55} & \ding{55} \\
    & CodeX~\cite{codex} & GPT-5~\cite{gpt5}            & \checkmark    & \ding{55} \\
    & Qwen Code~\cite{qwencode} & Qwen3-Coder-480B-A3B~\cite{qwen3technicalreport}  & \checkmark    & \checkmark \\ \midrule
    EvalAgent & PRDJudge  & Qwen3-Coder-30B-A3B~\cite{qwen3technicalreport}   &  \checkmark    & \checkmark  \\
\bottomrule
\end{tabular}
}
\end{table}

We evaluate two categories of code agents in our experiments: 
\begin{enumerate}[leftmargin=10pt]
    \item \textbf{Minimal Agents}\footnote{Code is available in https://github.com/AGI-Eval-Official/Minimal-CodeAgent.}, implemented using the Agent Development Kit (ADK). These agents are equipped with essential tools for file manipulation, bash scripting, and Python execution. We integrate state-of-the-art LLMs (Qwen3-Coder-480B-A3B, GPT-5.2, Claude-4.5-Sonnet, Gemini-3-Pro, Kimi-K2-Instruct-0905, DeepSeek-v3.2, GLM-4.7, and Minimax-M2) to systematically assess their core capabilities in strategy implementation. Hereafter we denote the models as Qwen3-Coder, GPT-5.2, Claude-4.5, Gemini-3-Pro, Kimi-K2, DeepSeek-v3.2, GLM-4.7, and Minimax-M2 respectively for simplicity.

    \item \textbf{Commercial Code Agents}, including advanced CLI-based agents (Claude Code, CodeX, Gemini CLI, and Qwen Code). These agents offer enhanced integration with command-line interfaces and external tools, representing the current state-of-the-art in code agent development.
\end{enumerate}

Each commercial agent and minimal agent utilizes a corresponding backbone LLM. Commercial agents typically benefit from vendor-specific fine-tuning and optimization, resulting in more stable and robust performance within their respective frameworks. While minimal agents are designed to facilitate fair comparison across different LLMs, commercial agents provide an assessment of the latest advancements in agent capabilities. The mapping between agent frameworks and backbone models is summarized in Table~\ref{tab:agent_models}.

\subsection{Experimental Setups}
For code agents, we select the latest model from each family that is compatible with the agent framework. For all LLMs used in our work, we set the temperature to be $0.1$, and the max tokens identical to their official APIs' setting. We set top-p to $1.0$, top-k to $100$, and presence penalty to be default to the API. Each code agent is executed using
a Python virtual environment that contains necessary and useful packages for the agents. If the code agent need other packages, we allow it use \texttt{pip} to install them. 

For PRDJudge training, we adopt Qwen3-Coder-30B~\cite{qwen3technicalreport} as our base model, balancing strong coding capabilities with deployment efficiency. To optimize training costs while maintaining performance, we employ Low-Rank Adaptation (LoRA) for parameter-efficient fine-tuning. The LoRA rank is set to 16. The model is trained for 1 epoch using a learning rate of $2.0 \times 10^{-4}$ and a warmup ratio of 0.03. The fine-tuning process is conducted on a cluster of 8 NVIDIA H800 GPUs.

\subsection{PRDJudge Evaluation}
\subsubsection{Evaluation Setup}
To evaluate the reliability and generalizability of PRDJudge, we construct two complementary test sets with full human annotation. (1) \textbf{In-Domain Test Set.} We collect code submissions generated by the same 11 code agents whose outputs were scored by the judge during training data construction. For each agent, we randomly sample 2 repositories (different from those used in training), yielding a total of 513 scoring metrics for human annotation. This test set is designed to assess PRDJudge's alignment with human judgment under familiar agent and repository distributions.  (2) \textbf{Out-of-Domain Test Set.} To evaluate the generalization capability of PRDJudge beyond the training distribution, we additionally construct an out-of-domain test set, where the code agents whose outputs are evaluated did not appear in the training data. This set comprises 26 repositories spanning diverse domains, resulting in a total of 605 scoring metrics. This setting more closely reflects real-world deployment scenarios, where PRDJudge may encounter previously unseen agent behaviors and implementation styles.

All annotations undergo a rigorous two-round labeling process to ensure quality and consistency. Specifically, each metric is independently annotated by two annotators. For cases where both annotators reach the same score, we randomly sample 10\% of such instances for a third-round quality check by a dedicated inspector. For cases where the two annotators disagree, the inspector reviews both annotations and provides a final adjudicated score, taking into account the reasoning of both parties. This multi-stage annotation pipeline ensures high inter-annotator reliability and minimizes subjective bias in the ground-truth labels.

\subsubsection{Evaluation Metric}
To assess the reliability of PRDJudge against human standards, we employ the {Human Alignment Rate (HAR)} as our primary evaluation metric. HAR calculates the exact match accuracy between the scores predicted by the evaluation model and the human-annotated ground truth. It intuitively reflects the overall correctness and reliability of the judge.

% To comprehensively assess the reliability of PRDJudge against human standards, we employ two primary evaluation metrics: (1) \textbf{Human Alignment Rate (HAR):} This metric calculates the exact match accuracy between the scores predicted by the evaluation agent and the human-annotated ground truth. It reflects the overall correctness of the judge.
% (2) \textbf{Quadratic Weighted Kappa (QWK):} Since our scoring system is ordinal (0, 1, 2), treating all misclassifications equally is suboptimal. For instance, misjudging completely broken code (Score 0) as perfect (Score 2) is a much more severe error than misjudging it as partially correct (Score 1). QWK addresses this by applying a quadratic penalty to the distance between the predicted and true scores, providing a more robust measure of alignment severity.

\begin{table}[t]
\caption{Performance comparison of PRDJudge and baseline models on the In-Domain and Out-of-Domain test sets. The accuracy is measured by the Human Alignment Rate (HAR, \%). Avg. Token and Avg. Time (in seconds) denote the average context length and inference time required per metric evaluation.}
\label{tab:PRDJudge}
\setlength{\tabcolsep}{1mm}
\renewcommand\arraystretch{1.0}
\resizebox{\linewidth}{!}{%
\begin{tabular}{lcccc}
\toprule
 \textbf{Model} & In-Domain & Out-of-Domain & Avg. Token & Avg.  Time \\ \midrule
 Human & 95.83 & 95.83 & - & > 600 \\ 
\midrule
  Qwen3-Coder (480B)     & 90.32 & 87.91 & 91,985 & 114.96   \\
  Claude-4.5      & 90.64        & 88.10         &  166,386 & 174.64 \\
  GPT-5.2     & 89.76             & 87.09        &  33,182 & 133.96 \\
  Qwen3-Coder (30B) & 70.52           &    55.46     & 151,411& 162.34 \\ 
  PRDJudge (ours) & \textbf{91.75} & \textbf{92.69} & 103,712  & 107.85\\\bottomrule

\end{tabular}
}
\end{table}

 \subsubsection{Human Alignment}
Table \ref{tab:PRDJudge} presents the evaluation results against human annotations, leading to three key conclusions:  (1) \textbf{PRDJudge achieves state-of-the-art human alignment.} PRDJudge outperforms all baseline models, including massive proprietary LLMs, and approaches the human inter-annotator agreement upper bound. The massive improvement over its base model (Qwen3-Coder 30B) highlights the necessity of specialized training for complex, long-context PRD-based evaluation. 
(2) \textbf{PRDJudge exhibits exceptional out-of-domain generalizability.} Zero-shot baseline models experience a noticeable performance drop on the out-of-domain test set, indicating this set is intrinsically harder (likely due to complex edge cases from unseen, stronger agents). Despite this challenge, PRDJudge maintains highly stable performance. This proves it genuinely internalizes the underlying evaluation criteria rather than merely overfitting to the training agents' specific error patterns. (3) \textbf{PRDJudge offers superior computational efficiency.} Alongside the highest accuracy, PRDJudge records the lowest average inference time. Compared to its base model, it significantly reduces token consumption and inference time, suggesting that specialized training enables more concise, targeted reasoning. This makes PRDJudge a highly practical and scalable solution for real-world automated evaluation.

\subsubsection{Stability Analysis}
To evaluate scoring stability, we conduct three independent evaluation runs for each metric and measure the Unanimous Agreement Rate (UAR) and Pairwise Agreement Rate (PAR). Detailed results are provided in Appendix \ref{app:stability}. PRDJudge demonstrates the highest stability among all evaluated models, achieving a UAR of 94.19\% and a PAR of 96.07\%. The superior consistency indicates that our specialized fine-tuning helps PRDJudge internalize a deterministic evaluation mapping from PRDs to code, significantly mitigating the reasoning variance and generation randomness typically observed in general-purpose LLMs.

\subsubsection{Case Study}
To intuitively understand why PRDJudge achieves superior performance and efficiency, we trace and analyze the tool-use trajectories of different models during the evaluation process. The complete tool usage distribution and case study is provided in Appendix~\ref{app:tool_usage}. First, PRDJudge follows a logical explore-then-evaluate workflow, prioritizing repository exploration and file reading to understand code logic before making judgments. In contrast, GPT-5.2 exhibits a blind action pattern, characterized by excessive command execution and file writing with minimal code inspection. Second, PRDJudge demonstrates high precision. While its base model (Qwen3-30B) suffers from tool spamming due to convoluted reasoning, PRDJudge requires significantly fewer, more targeted calls than even Claude-4.5. This confirms that specialized training enables the model to optimally gather evidence rather than merely increasing tool-use frequency.

\subsection{Code Agent Evaluation}
\subsubsection{Overall Results}
Table~\ref{tab:mainresults} presents agent performances across two phases: Round 1 (DEV) assesses zero-shot project implementation based on PRDs, while Round 2 (DEBUG) evaluates iterative refinement using PRDJudge feedback. Our main findings are: \textbf{(1) Foundation models dictate zero-shot development ceilings.} Frontier models (e.g., Claude-4.5, GPT-5.2) consistently dominate the DEV phase, indicating a strong correlation between intrinsic LLM reasoning capacity and the ability to generate project-level scaffolds from scratch. \textbf{(2) Commercial frameworks constrain initial generation but excel in iterative debugging.} Minimal agents generally outperform their Commercial counterparts in DEV, suggesting that complex tool-use guardrails in commercial frameworks restrict zero-shot generation. Conversely, commercial agents demonstrate superior debugging capabilities, effectively utilizing feedback for targeted fixes. \textbf{(3) Unconstrained models risk performance degradation during long-context debugging.} High-performing Minimal agents (e.g., Claude-4.5, DeepSeek-V3.2) experience substantial drops in the DEBUG phase. Without robust agentic scaffolding to manage global states, attempting localized modifications frequently breaks previously functioning modules, underscoring the necessity of structured designs for safe code maintenance. \textbf{(4) LLMs exhibit divergent generation and refinement profiles.} Within Minimal agents, we observe three archetypes: (i) strong zero-shot generation but high variance in context maintenance, causing debugging regressions (Claude-4.5, DeepSeek-V3.2); (ii) limited initial generation but strong instruction-following for localized fixes, yielding significant improvements (Kimi-K2, Gemini-3-Pro); and (iii) consistent performance across both phases (GPT-5.2, Qwen3-Coder). This suggests future agent designs must adapt to the specific reasoning characteristics of their backbone models.

\begin{table}[t]
\caption{Average pass rate of code agents on PRDBench (in \%). The best results are highlighted in \textbf{bold}, and the second-best results are \underline{underlined}.}
\label{tab:mainresults}
\setlength{\tabcolsep}{2mm}
\renewcommand\arraystretch{0.9}
\resizebox{\linewidth}{!}{%
\begin{tabular}{llccc}
\toprule
\textbf{Agent Type} & \textbf{Agent} & \multicolumn{1}{c}{\textbf{DEV}$\uparrow$} & \multicolumn{1}{c}{\textbf{DEBUG}$\uparrow$} & \textbf{Enhance}$\uparrow$   \\ \midrule
\multirow{8}{*}{\textbf{Minimal}} 
 & GPT-5.2       & \underline{62.49} & \underline{69.00} &  +6.51 \\
 & Claude-4.5      & \textbf{69.19}   &     56.40    &  -12.79  \\
 & Gemini-3-Pro      & 22.76          &  27.28     &  +4.52 \\
 & Qwen3-Coder & 43.84         &    48.29       & +4.45 \\
 & Kimi-K2 & 20.52 &  36.17& +15.65 \\ 
 & DeepSeek-V3.2 & 40.11  & 24.80 & -15.31 \\ 
 & GLM-4.7 &  38.39 & 35.33 &  -3.06\\ 
 & Minimax-M2 & 17.60  & 24.75 &  +7.15\\ 
 \midrule
\multirow{4}{*}{\textbf{Commercial}} 
 & CodeX       &  62.09 &  65.02 &  +2.97\\
 & Claude Code & 56.65            &    \textbf{70.25}    & +13.60 \\
 & Gemini CLI  & 11.29          &     21.51       &  +10.22  \\
 & Qwen Code   & 39.91            &    35.95     & -3.96  \\ \bottomrule
\end{tabular}
% \vspace{-15pt}
}
\end{table}

\begin{figure*}[t]
    \centering
    \includegraphics[width=\linewidth]{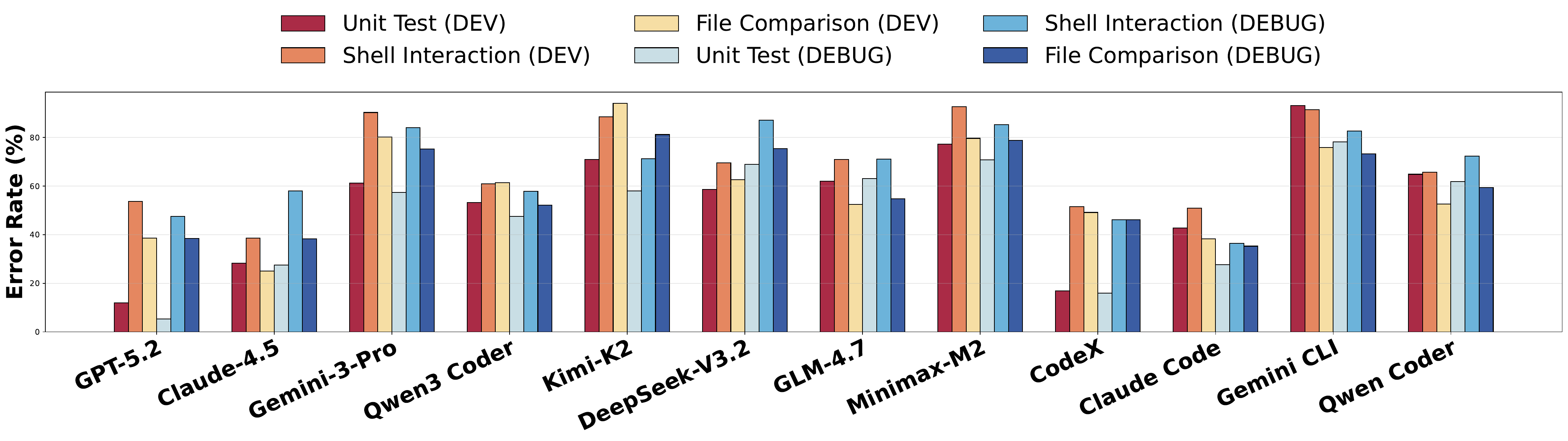}
    \vspace{-10pt}
    \caption{Error rates of code agents on different types of test cases.}
    \vspace{-5pt}
    \label{fig:proportion}
\end{figure*}

\subsubsection{Cost of Code Agents}
Appendix~\ref{app:costofagent} details the computational expenditures of the evaluated LLM-based code agents across two experimental iterations, revealing distinct operational patterns. First, the generation and maintenance profiles diverge significantly across agent configurations. The execution of minimal LLM agents is characterized by high-volume initial code generation, followed by high volatility during the debugging phase, where modifications vary from minor syntax corrections to extensive logic restructuring within the context window. Conversely, commercial LLM frameworks operate under highly constrained generation parameters. Their maintenance phase is defined by localized modifications, which restricts code alterations to a stable and minimal scope regardless of task complexity. Second, the underlying drivers of computational overhead differ fundamentally. For minimal agents, token consumption and inference latency scale directly with the volume of generated code. In contrast, commercial agents incur substantial token expenditures and latency that remain largely decoupled from the actual volume of code modification. This resource distribution reflects an architectural reliance on complex tool-use reasoning, continuous environment synchronization, and iterative state management rather than baseline zero-shot drafting speed. 

\subsubsection{Error Analysis of Code Agent}
Figure~\ref{fig:proportion} shows the error rates of code agents across different types of test cases. The error rate reflects the proportion of failed test cases within each category for the code generated by the agents. This comparison highlights the strengths and weaknesses of various agents when handling diverse evaluation scenarios. We observe two main findings from the results. First, there is no single category of test cases that is consistently easy for code agents. The error rates are relatively uniform across the three types of test cases, indicating that PRDBench’s test case design is well-balanced and effectively covers a broad spectrum of code implementation scenarios. Second, unit test cases are noticeably more challenging to debug compared to the other two types. Most code agents fail to resolve these issues even after the debugging phase. This is primarily because unit test cases require agents to read and understand the test function code before making corrections, whereas the other two test types only require agents to compare actual and expected outputs, which lowers the reasoning complexity.

\subsubsection{Free Development}
\begin{figure}[t]
    \centering
    \includegraphics[width=0.9\linewidth]{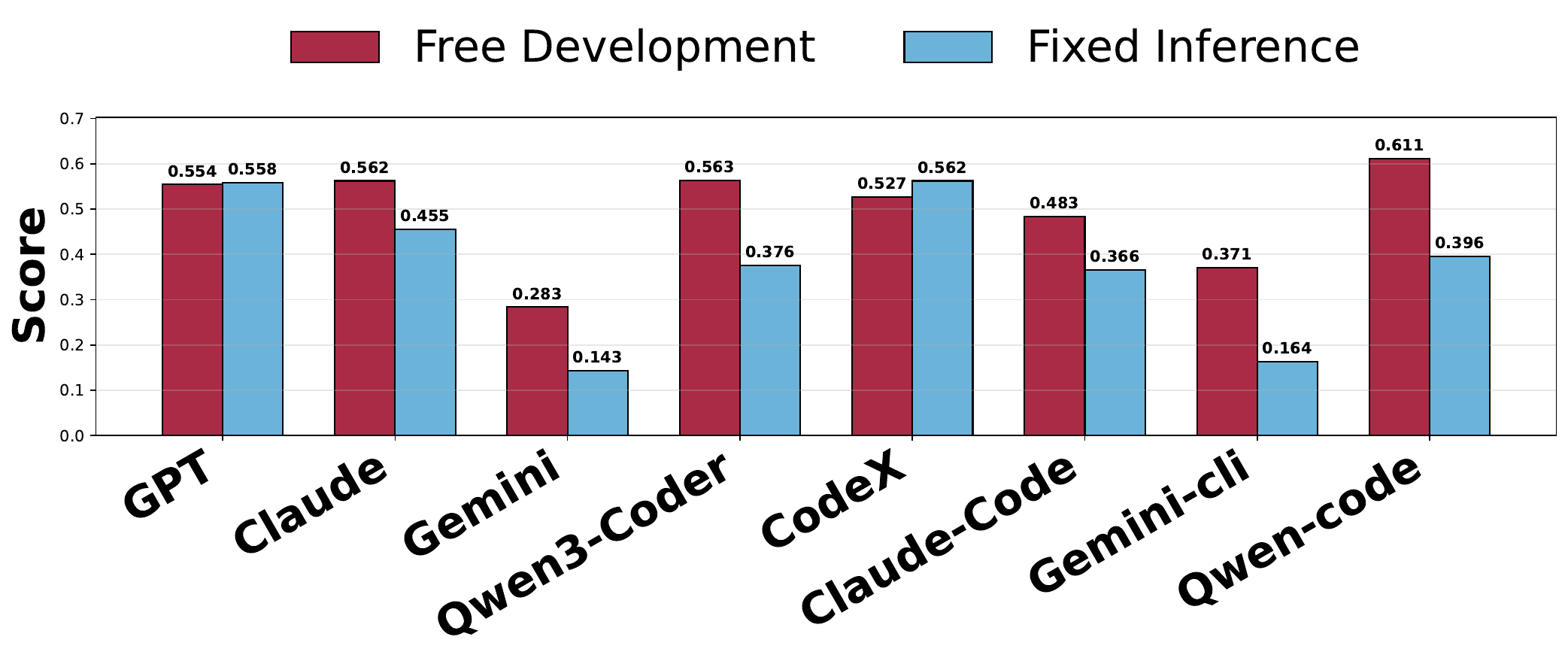}
\vspace{-10pt}
    \caption{Comparison of code agent scores under fixed inference and free development modes on PRDBench.}
    % \vspace{-10pt}
    \label{fig:freedev}
\end{figure}

In real-world scenarios, developers often build from scratch based solely on a PRD. Thanks to its scaffold-removal design, PRDBench inherently supports this unconstrained free development setting. To explore this, we prompted PRDJudge to dynamically adapt test interfaces to the agents' free-form code.

Figure~\ref{fig:freedev} compares agent performance across both modes. While relative rankings remain stable, the performance variance decreases significantly under free development. This highlights a key confounding variable: it is difficult to disentangle agent logic errors from PRDJudge's failures in dynamically rewriting test interfaces. Due to this evaluation noise, we conclude that Fixed Inference remains the most reliable benchmarking setting at present. The free development mode serves as a proof-of-concept, underscoring the inherent difficulties of unconstrained evaluation that we aim to address in future work.

\section{Conclusion}
In this paper, we introduce PRDBench, a comprehensive benchmark for evaluating repository-level code agents against Product Requirement Documents (PRDs). To address the scalability limitations of human evaluation and the inherent inefficiencies of general-purpose LLM judges, we propose PRDJudge, a specialized evaluator fine-tuned for code-to-PRD alignment. PRDJudge achieves state-of-the-art human alignment and exceptional out-of-domain generalizability. Trajectory analysis demonstrates that PRDJudge adopts a systematic verification strategy, effectively mitigating the redundant tool invocations and execution-heavy inefficiencies prevalent in general LLMs, thereby providing a robust evaluation infrastructure. Leveraging this framework, our extensive evaluations on PRDBench reveal critical insights into current code agents. We observe that while the intrinsic capabilities of foundation models determine the ceiling for zero-shot development, commercial frameworks constrain initial generation but significantly enhance iterative debugging. Furthermore, unconstrained LLMs frequently introduce regressions during long-context refinement, and system-level interactions pose far greater challenges than isolated unit tests. Ultimately, by shifting the focus from isolated functional correctness to comprehensive PRD alignment, this work establishes a scalable and robust evaluation paradigm, providing a critical foundation for the future development and assessment of code agents.

\textbf{Future Work.} While PRDBench excels in fixed-interface settings, evaluating unconstrained Free Development introduces noise when dynamically adapting test scripts to unpredictable code. Future work will focus on enhancing PRDJudge's dynamic interface-mapping capabilities to achieve reliable automated evaluation in fully unconstrained scenarios.

\begin{acks}
The Shanghai Jiao Tong University team is partially supported by National Natural Science Foundation of China (62322603, 62177033, 62502310). We extend our sincere gratitude to Shuang Zhou, \begin{CJK*}{UTF8}{gbsn}Yuan Zhang (张元), Yuan Zhang (张源)\end{CJK*},  Ying Xie, Hao Zheng, Yan Wang, Zhiqiang Li, Cheng Li, Bangpeng Wei and Huaying Xia for their contributions to data annotation.
\end{acks}

%%%%%%%%%%%%%%%%%%%%%%%%%%%%%%%%%%%%%%%%%%%%%%%%%%%%%%%%%%%%%%%%%%%%%%%%

%%% The next two lines define, first, the bibliography style to be 
%%% applied, and, second, the bibliography file to be used.

\bibliographystyle{ACM-Reference-Format}
\balance
\bibliography{sample}

@article{wei2025developers,
  title={How Do Developers Structure Unit Test Cases? An Empirical Analysis of the AAA Pattern in Open Source Projects},
  author={Wei, Chenhao and Xiao, Lu and Yu, Tingting and Wong, Sunny and Clune, Abigail},
  journal={IEEE Transactions on Software Engineering},
  year={2025},
  publisher={IEEE}
}

@article{starace2025paperbench,
  title={PaperBench: Evaluating AI's Ability to Replicate AI Research},
  author={Starace, Giulio and Jaffe, Oliver and Sherburn, Dane and Aung, James and Chan, Jun Shern and Maksin, Leon and Dias, Rachel and Mays, Evan and Kinsella, Benjamin and Thompson, Wyatt and others},
  journal={arXiv preprint arXiv:2504.01848},
  year={2025}
}

@article{zhuge2024agent,
  title={Agent-as-a-judge: Evaluate agents with agents},
  author={Zhuge, Mingchen and Zhao, Changsheng and Ashley, Dylan and Wang, Wenyi and Khizbullin, Dmitrii and Xiong, Yunyang and Liu, Zechun and Chang, Ernie and Krishnamoorthi, Raghuraman and Tian, Yuandong and others},
  journal={arXiv preprint arXiv:2410.10934},
  year={2024}
}

@article{chan2024mlebench,
  title={Mle-bench: Evaluating machine learning agents on machine learning engineering},
  author={Chan, Jun Shern and Chowdhury, Neil and Jaffe, Oliver and Aung, James and Sherburn, Dane and Mays, Evan and Starace, Giulio and Liu, Kevin and Maksin, Leon and Patwardhan, Tejal and others},
  journal={arXiv preprint arXiv:2410.07095},
  year={2024}
}

@article{jimenez2023swe,
  title={Swe-bench: Can language models resolve real-world github issues?},
  author={Jimenez, Carlos E and Yang, John and Wettig, Alexander and Yao, Shunyu and Pei, Kexin and Press, Ofir and Narasimhan, Karthik},
  journal={arXiv preprint arXiv:2310.06770},
  year={2023}
}

@misc{anysphere2025cursoragent,
  author = {Anysphere Inc.},
  title = {Cursor: The AI powered Code Editor},
  year = {2025},
  howpublished = {\url{https://cursor.com/}},
  note = {AI assistant code editor announcement, Anysphere blog}
}

@misc{wang2025codevisionaryagentbasedframeworkevaluating,
      title={CodeVisionary: An Agent-based Framework for Evaluating Large Language Models in Code Generation}, 
      author={Xinchen Wang and Pengfei Gao and Chao Peng and Ruida Hu and Cuiyun Gao},
      year={2025},
      eprint={2504.13472},
      archivePrefix={arXiv},
      primaryClass={cs.SE},
      url={https://arxiv.org/abs/2504.13472}, 
}

@misc{mullen2025geminicli,
  author = {Taylor Mullen and Ryan J. Salva},
  title = {Gemini CLI: Your Open Source AI Agent},
  year = {2025},
  howpublished = {\url{https://blog.google/technology/developers/introducing-gemini-cli-open-source-ai-agent/}},
  note = {Google Developers Blog, Jun 2025}
}

@misc{vergopoulos2025automatedbenchmarkgenerationrepositorylevel,
      title={Automated Benchmark Generation for Repository-Level Coding Tasks}, 
      author={Konstantinos Vergopoulos and Mark Niklas Müller and Martin Vechev},
      year={2025},
      eprint={2503.07701},
      archivePrefix={arXiv},
      primaryClass={cs.SE},
      url={https://arxiv.org/abs/2503.07701}, 
}

@article{fu2025corecodebench,
  title={CoreCodeBench: A Configurable Multi-Scenario Repository-Level Benchmark},
  author={Fu, Lingyue and Guan, Hao and Zhang, Bolun and Yuan, Haowei and Zhu, Yaoming and Xu, Jun and Wang, Zongyu and Qiu, Lin and Cai, Xunliang and Cao, Xuezhi and others},
  journal={arXiv preprint arXiv:2507.05281},
  year={2025}
}

@misc{daghighfarsoodeh2025deepbenchdeeplearningbenchmark,
      title={Deep-Bench: Deep Learning Benchmark Dataset for Code Generation}, 
      author={Alireza Daghighfarsoodeh and Chung-Yu Wang and Hamed Taherkhani and Melika Sepidband and Mohammad Abdollahi and Hadi Hemmati and Hung Viet Pham},
      year={2025},
      eprint={2502.18726},
      archivePrefix={arXiv},
      primaryClass={cs.SE},
      url={https://arxiv.org/abs/2502.18726}, 
}

@misc{liu2025projectevalbenchmarkprogrammingagents,
      title={ProjectEval: A Benchmark for Programming Agents Automated Evaluation on Project-Level Code Generation}, 
      author={Kaiyuan Liu and Youcheng Pan and Yang Xiang and Daojing He and Jing Li and Yexing Du and Tianrun Gao},
      year={2025},
      eprint={2503.07010},
      archivePrefix={arXiv},
      primaryClass={cs.SE},
      url={https://arxiv.org/abs/2503.07010}, 
}

@misc{huang2024dacodeagentdatascience,
      title={DA-Code: Agent Data Science Code Generation Benchmark for Large Language Models}, 
      author={Yiming Huang and Jianwen Luo and Yan Yu and Yitong Zhang and Fangyu Lei and Yifan Wei and Shizhu He and Lifu Huang and Xiao Liu and Jun Zhao and Kang Liu},
      year={2024},
      eprint={2410.07331},
      archivePrefix={arXiv},
      primaryClass={cs.CL},
      url={https://arxiv.org/abs/2410.07331}, 
}

@misc{xiang2025scireplicatebenchbenchmarkingllmsagentdriven,
      title={SciReplicate-Bench: Benchmarking LLMs in Agent-driven Algorithmic Reproduction from Research Papers}, 
      author={Yanzheng Xiang and Hanqi Yan and Shuyin Ouyang and Lin Gui and Yulan He},
      year={2025},
      eprint={2504.00255},
      archivePrefix={arXiv},
      primaryClass={cs.CL},
      url={https://arxiv.org/abs/2504.00255}, 
}

@misc{li2024promptinglargelanguagemodels,
      title={Prompting Large Language Models to Tackle the Full Software Development Lifecycle: A Case Study}, 
      author={Bowen Li and Wenhan Wu and Ziwei Tang and Lin Shi and John Yang and Jinyang Li and Shunyu Yao and Chen Qian and Binyuan Hui and Qicheng Zhang and Zhiyin Yu and He Du and Ping Yang and Dahua Lin and Chao Peng and Kai Chen},
      year={2024},
      eprint={2403.08604},
      archivePrefix={arXiv},
      primaryClass={cs.CL},
      url={https://arxiv.org/abs/2403.08604}, 
}

@misc{qwencode,
  author = {Qwen Team},
  title = {Qwen Code brings the capabilities of advanced code models to your terminal in an interactive Read-Eval-Print Loop (REPL) environment},
  year = {2025},
  note = {Accessed: 2025-07-22},
  howpublished = {\url{https://qwenlm.github.io/qwen-code-docs/en/}}
}

@misc{gpt5,
  author = {OpenAI},
  title = {Introducing GPT-5},
  year = {2025},
  note = {Accessed: 2025-08-07},
  howpublished = {\url{https://openai.com/index/introducing-gpt-5/}}
}

@misc{gpt5.2,
   author = {OpenAI},
  title = {Introducing GPT-5.2},
  year = {2025},
  note = {Accessed: 2025-12-11},
    howpublished={\url{https://openai.com/zh-Hans-CN/index/introducing-gpt-5-2/}}
}

@misc{anthropic2025claude37,
  author = {Anthropic},
  title = {Claude 3.7 Sonnet and Claude Code},
  year = {2025},
  note = {Accessed: 2025-02-25},
  howpublished = {\url{https://www.anthropic.com/news/claude-3-7-sonnet}}
}

@misc{kimiteam2025kimik2openagentic,
      title={Kimi K2: Open Agentic Intelligence}, 
      author={Kimi Team},
      year={2025},
      eprint={2507.20534},
      archivePrefix={arXiv},
      primaryClass={cs.LG},
      url={https://arxiv.org/abs/2507.20534}, 
}

@misc{anthropicclaude45,
title={Introducing Claude Sonnet 4.5}, 
      author={Anthropic},
      year = {2025},
  note = {Accessed: 2025-09-25},
  howpublished = {\url{https://www.anthropic.com/news/claude-sonnet-4-5}}
}

@misc{5team2025glm45agenticreasoningcoding,
      title={GLM-4.5: Agentic, Reasoning, and Coding (ARC) Foundation Models}, 
      author={GLM Team},
      year={2025},
      eprint={2508.06471},
      archivePrefix={arXiv},
      primaryClass={cs.CL},
      url={https://arxiv.org/abs/2508.06471}, 
}

@misc{gemini25,
      title={Gemini 2.5: Pushing the Frontier with Advanced Reasoning, Multimodality, Long Context, and Next Generation Agentic Capabilities}, 
      author={Gemini Team},
      year={2025},
      eprint={2507.06261},
      archivePrefix={arXiv},
      primaryClass={cs.CL},
      url={https://arxiv.org/abs/2507.06261}, 
}

@misc{gemini3,
title={Best for complex tasks and bringing creative concepts to life},
author={Gemini Team},
year={2025},
howpublised={\url{https://deepmind.google/models/gemini/pro/}}
}

@misc{minimaxm2,
    title={MiniMax M2 \& Agent: Ingenious in Simplicity},
author={Minimax Team},
year={2025},
   howpublished ={\url{https://www.minimax.io/news/minimax-m2}} 
}

@misc{codex,
    author={OpenAI},
title= {CodeX CLI},
year={2025},
note = {Accessed: 2025-05-16},
howpublished={\url{https://github.com/openai/codex}}
}

@misc{qwen3technicalreport,
      title={Qwen3 Technical Report}, 
      author={Qwen Team},
      year={2025},
      eprint={2505.09388},
      archivePrefix={arXiv},
      primaryClass={cs.CL},
      url={https://arxiv.org/abs/2505.09388}, 
}

@inproceedings{chen2024mllm,
  title={Mllm-as-a-judge: Assessing multimodal llm-as-a-judge with vision-language benchmark},
  author={Chen, Dongping and Chen, Ruoxi and Zhang, Shilin and Wang, Yaochen and Liu, Yinuo and Zhou, Huichi and Zhang, Qihui and Wan, Yao and Zhou, Pan and Sun, Lichao},
  booktitle={Forty-first International Conference on Machine Learning},
  year={2024}
}

@misc{weyssow2024codeultrafeedbackllmasajudgedatasetaligning,
      title={CodeUltraFeedback: An LLM-as-a-Judge Dataset for Aligning Large Language Models to Coding Preferences}, 
      author={Martin Weyssow and Aton Kamanda and Xin Zhou and Houari Sahraoui},
      year={2024},
      eprint={2403.09032},
      archivePrefix={arXiv},
      primaryClass={cs.SE},
      url={https://arxiv.org/abs/2403.09032}, 
}

@article{ho2025llm,
  title={LLM-as-a-Judge: Reassessing the Performance of LLMs in Extractive QA},
  author={Ho, Xanh and Huang, Jiahao and Boudin, Florian and Aizawa, Akiko},
  journal={arXiv preprint arXiv:2504.11972},
  year={2025}
}

@misc{wu2024repomasterevalevaluatingcodecompletion,
      title={RepoMasterEval: Evaluating Code Completion via Real-World Repositories}, 
      author={Qinyun Wu and Chao Peng and Pengfei Gao and Ruida Hu and Haoyu Gan and Bo Jiang and Jinhe Tang and Zhiwen Deng and Zhanming Guan and Cuiyun Gao and Xia Liu and Ping Yang},
      year={2024},
      eprint={2408.03519},
      archivePrefix={arXiv},
      primaryClass={cs.SE},
      url={https://arxiv.org/abs/2408.03519}, 
}

@misc{hai2025impactscontextsrepositorylevelcode,
      title={On the Impacts of Contexts on Repository-Level Code Generation}, 
      author={Nam Le Hai and Dung Manh Nguyen and Nghi D. Q. Bui},
      year={2025},
      eprint={2406.11927},
      archivePrefix={arXiv},
      primaryClass={cs.SE},
      url={https://arxiv.org/abs/2406.11927}, 
}

@misc{jimenez2024swebenchlanguagemodelsresolve,
      title={SWE-bench: Can Language Models Resolve Real-World GitHub Issues?}, 
      author={Carlos E. Jimenez and John Yang and Alexander Wettig and Shunyu Yao and Kexin Pei and Ofir Press and Karthik Narasimhan},
      year={2024},
      eprint={2310.06770},
      archivePrefix={arXiv},
      primaryClass={cs.CL},
      url={https://arxiv.org/abs/2310.06770}, 
}

@misc{pan2024codevbenchllmsunderstanddevelopercentric,
      title={Codev-Bench: How Do LLMs Understand Developer-Centric Code Completion?}, 
      author={Zhenyu Pan and Rongyu Cao and Yongchang Cao and Yingwei Ma and Binhua Li and Fei Huang and Han Liu and Yongbin Li},
      year={2024},
      eprint={2410.01353},
      archivePrefix={arXiv},
      primaryClass={cs.SE},
      url={https://arxiv.org/abs/2410.01353}, 
}

@misc{yang2024execrepobenchmultilevelexecutablecode,
      title={ExecRepoBench: Multi-level Executable Code Completion Evaluation}, 
      author={Jian Yang and Jiajun Zhang and Jiaxi Yang and Ke Jin and Lei Zhang and Qiyao Peng and Ken Deng and Yibo Miao and Tianyu Liu and Zeyu Cui and Binyuan Hui and Junyang Lin},
      year={2024},
      eprint={2412.11990},
      archivePrefix={arXiv},
      primaryClass={cs.CL},
      url={https://arxiv.org/abs/2412.11990}, 
}

@misc{li2024evocodebenchevolvingcodegeneration,
      title={EvoCodeBench: An Evolving Code Generation Benchmark Aligned with Real-World Code Repositories}, 
      author={Jia Li and Ge Li and Xuanming Zhang and Yihong Dong and Zhi Jin},
      year={2024},
      eprint={2404.00599},
      archivePrefix={arXiv},
      primaryClass={cs.CL},
      url={https://arxiv.org/abs/2404.00599}, 
}

@misc{zhuo2025bigcodebenchbenchmarkingcodegeneration,
      title={BigCodeBench: Benchmarking Code Generation with Diverse Function Calls and Complex Instructions}, 
      author={Terry Yue Zhuo and Minh Chien Vu and Jenny Chim and Han Hu and Wenhao Yu and Ratnadira Widyasari and Imam Nur Bani Yusuf and Haolan Zhan and Junda He and Indraneil Paul and Simon Brunner and Chen Gong and Thong Hoang and Armel Randy Zebaze and Xiaoheng Hong and Wen-Ding Li and Jean Kaddour and Ming Xu and Zhihan Zhang and Prateek Yadav and Naman Jain and Alex Gu and Zhoujun Cheng and Jiawei Liu and Qian Liu and Zijian Wang and Binyuan Hui and Niklas Muennighoff and David Lo and Daniel Fried and Xiaoning Du and Harm de Vries and Leandro Von Werra},
      year={2025},
      eprint={2406.15877},
      archivePrefix={arXiv},
      primaryClass={cs.SE},
      url={https://arxiv.org/abs/2406.15877}, 
}

@misc{peng2024humanevalxlmultilingualcodegeneration,
      title={HumanEval-XL: A Multilingual Code Generation Benchmark for Cross-lingual Natural Language Generalization}, 
      author={Qiwei Peng and Yekun Chai and Xuhong Li},
      year={2024},
      eprint={2402.16694},
      archivePrefix={arXiv},
      primaryClass={cs.CL},
      url={https://arxiv.org/abs/2402.16694}, 
}

@misc{together2024finetuning,
  author = {Together AI},
  title = {Fine-tuning Open LLM Judges to Outperform GPT-4},
  year = {2024},
  url = {https://www.together.ai/blog/fine-tuning-open-llm-judges-to-outperform-gpt-5-2},
  note = {Accessed: 2024-05-20}
}

@misc{huang2025empiricalstudyllmasajudgellm,
      title={An Empirical Study of LLM-as-a-Judge for LLM Evaluation: Fine-tuned Judge Model is not a General Substitute for GPT-4}, 
      author={Hui Huang and Xingyuan Bu and Hongli Zhou and Yingqi Qu and Jing Liu and Muyun Yang and Bing Xu and Tiejun Zhao},
      year={2025},
      eprint={2403.02839},
      archivePrefix={arXiv},
      primaryClass={cs.CL},
      url={https://arxiv.org/abs/2403.02839}, 
}

@misc{li2023generativejudgeevaluatingalignment,
      title={Generative Judge for Evaluating Alignment}, 
      author={Junlong Li and Shichao Sun and Weizhe Yuan and Run-Ze Fan and Hai Zhao and Pengfei Liu},
      year={2023},
      eprint={2310.05470},
      archivePrefix={arXiv},
      primaryClass={cs.CL},
      url={https://arxiv.org/abs/2310.05470}, 
}

@misc{zhu2025judgelmfinetunedlargelanguage,
      title={JudgeLM: Fine-tuned Large Language Models are Scalable Judges}, 
      author={Lianghui Zhu and Xinggang Wang and Xinlong Wang},
      year={2025},
      eprint={2310.17631},
      archivePrefix={arXiv},
      primaryClass={cs.CL},
      url={https://arxiv.org/abs/2310.17631}, 
}

@misc{kim2024prometheusinducingfinegrainedevaluation,
      title={Prometheus: Inducing Fine-grained Evaluation Capability in Language Models}, 
      author={Seungone Kim and Jamin Shin and Yejin Cho and Joel Jang and Shayne Longpre and Hwaran Lee and Sangdoo Yun and Seongjin Shin and Sungdong Kim and James Thorne and Minjoon Seo},
      year={2024},
      eprint={2310.08491},
      archivePrefix={arXiv},
      primaryClass={cs.CL},
      url={https://arxiv.org/abs/2310.08491}, 
}

@misc{wang2024pandalmautomaticevaluationbenchmark,
      title={PandaLM: An Automatic Evaluation Benchmark for LLM Instruction Tuning Optimization}, 
      author={Yidong Wang and Zhuohao Yu and Zhengran Zeng and Linyi Yang and Cunxiang Wang and Hao Chen and Chaoya Jiang and Rui Xie and Jindong Wang and Xing Xie and Wei Ye and Shikun Zhang and Yue Zhang},
      year={2024},
      eprint={2306.05087},
      archivePrefix={arXiv},
      primaryClass={cs.CL},
      url={https://arxiv.org/abs/2306.05087}, 
}

%%%%%%%%%%%%%%%%%%%%%%%%%%%%%%%%%%%%%%%%%%%%%%%%%%%%%%%%%%%%%%%%%%%%%%%%

\appendix
\newpage
\appendix

\section{Recruitment and Compensation of Annotators}
\label{sec:background}

We employ a total of 9 annotators, including six full-time annotators and three part-time annotators. We list their educational background in Table~\ref{tab:background}.

Among them, F1 to F6 and P1 participants in PRDBench construction and each of them contribute at least 5 problems.

F1, F2, P2 and P3 participants in data examination. 

F1 to F6, P2 and P3 participants in human score annotation to evaluate and train PRDJudge. 

\begin{table*}[htb]
\caption{Background of annotators. }
\label{tab:background}
\begin{tabular}{l l l l c}
\hline
Type & ID & Major & Degree & Years of Experience as Developer \\
\hline
Full-time & F1 & Computer Science & Master & 10 \\
Full-time &  F2 & Software Engineering & Master & 6 \\
Full-time &  F3 & Applied Statistics & Master & 1 \\
Full-time &  F4 & Health Service and Management & Bachelor & 2 \\
Full-time &  F5 & Information Management & Bachelor & 4 \\
Full-time &  F6 & Economics and Trade Management & Bachelor & 6 \\
Part-time &  P1 & Computer Science & Bachelor & 9 \\
Part-time &  P2 & Computer Science & Senior Undergraduate & - \\
Part-time &  P3 & Computer Science & Senior Undergraduate & - \\
\hline
\end{tabular}
\end{table*}

All annotators receive daily wages~(part-time) or monthly salaries~(full-time) according to local labor regulations.

\section{Stability Analysis of PRDJudge}\label{app:stability}
In this section, we provide the detailed evaluation metrics and the complete results for the stability analysis discussed in the main text.

\subsection{Metrics Definition}
Due to the inherent probabilistic nature of Large Language Models, their generated evaluations can exhibit variance across different runs, especially when dealing with long-context reasoning tasks like PRD-based code evaluation. To rigorously quantify scoring stability, we conduct three independent evaluation runs for each metric and employ two criteria:

\begin{itemize}[leftmargin=10pt]
    \item \textbf{Unanimous Agreement Rate (UAR):} This metric calculates the percentage of evaluation instances where the model predicts the exact same score across all three independent runs. Let $s_1, s_2, s_3$ be the predicted scores in three runs. The UAR for a single metric is 1 if $s_1 = s_2 = s_3$, and 0 otherwise. It represents the strictest measure of consistency.
    
    \item \textbf{Pairwise Agreement Rate (PAR):} This metric computes the average agreement rate across all three possible pairs of the runs (i.e., Run 1 vs. Run 2, Run 2 vs. Run 3, and Run 1 vs. Run 3). It provides a more granular view of stability, accounting for partial agreements even when unanimous agreement is not reached.
\end{itemize}

\subsection{Results and Discussion}
As shown in Table \ref{tab:stability_appendix}, evaluating code based on complex PRDs is a highly challenging task that can induce reasoning fluctuations in general-purpose LLMs. Even state-of-the-art proprietary models like Claude-4.5 and GPT-5.2 exhibit noticeable variance, failing to reach unanimous agreement in approximately 7\% to 9\% of the cases. This variance often stems from the models attending to different constraints within the lengthy PRD context across different runs. In contrast, PRDJudge achieves the highest stability with a UAR of 94.19\% and a PAR of 96.07\%. Through specialized fine-tuning on domain-specific evaluation trajectories, PRDJudge has learned a more robust and deterministic mapping between PRD requirements and code implementations. This significantly mitigates the generation randomness, making PRDJudge a more reliable evaluator for automated pipelines where consistent feedback is critical.

\begin{table}[h]
\centering
\caption{Stability analysis of different models across three independent runs. UAR denotes Unanimous Agreement Rate, and PAR denotes Pairwise Agreement Rate. Higher values indicate better scoring stability.}
\label{tab:stability_appendix}
\setlength{\tabcolsep}{3mm}
\begin{tabular}{lcc}
\toprule
\textbf{Model} & \textbf{UAR (\%)} $\uparrow$ & \textbf{PAR (\%)} $\uparrow$ \\ \midrule
Qwen3-Coder (480B) &91.95 &
94.57 \\
Claude-4.5 & 93.07 &
95.32 \\
GPT-5.2 &91.22 &
94.14 \\
Qwen3-Coder (30B) & 70.92 &  79.42\\ \midrule
\textbf{PRDJudge (ours)} & \textbf{94.19} &
\textbf{96.07} \\ \bottomrule
\end{tabular}
\end{table}

\section{EvalAgent Analysis}
\label{app:tool_usage}

\subsection{Tool Use Distribution}
\begin{figure}[htb]
    \centering
    \includegraphics[width=\linewidth]{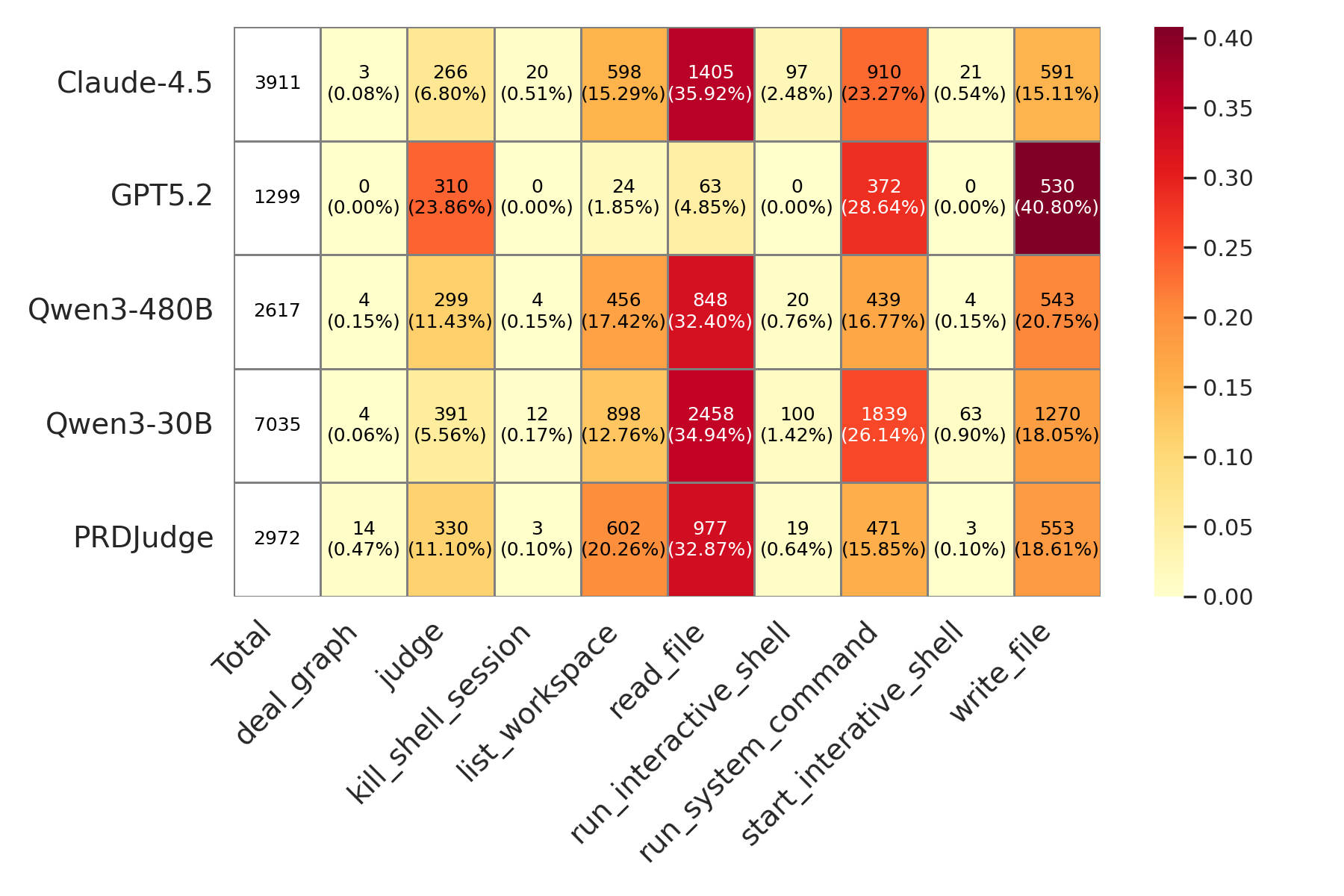}
    \caption{Tool Use Distribution of Agents. }
    \label{fig:tooluse}
\end{figure}

Figure \ref{fig:tooluse} illustrates the distribution and total frequency of tool calls made by different models during the evaluation process. This behavioral analysis provides deeper insights into the reasoning strategies adopted by each evaluator.

The heatmap highlights several critical anomalies in general-purpose LLMs when confronted with complex, repository-level evaluation tasks:
\begin{itemize}[leftmargin=10pt]
    \item \textbf{GPT-5.2's Skewed Trajectory:} GPT-5.2 exhibits a highly irregular pattern. It rarely uses exploration tools like \texttt{list workspace} (1.85\%) and \texttt{read file} (4.85\%). Instead, it heavily relies on \texttt{write file} (40.80\%) and \texttt{run system command} (28.64\%). This suggests that GPT-5.2 attempts to solve the evaluation by aggressively executing trial-and-error operations rather than systematically reading the codebase.
    \item \textbf{Qwen3-Coder (30B)'s Inefficiency:} The base model, Qwen3-30B, records an exorbitant 7,035 total tool calls. This massive volume indicates that the model frequently falls into reasoning loops or hallucinates incorrect tool parameters, forcing it to retry repeatedly without progressing toward a final judgment.
\end{itemize}

Conversely, {PRDJudge} demonstrates an optimal balance. It dedicates the highest proportion among all models to repository exploration (\texttt{list workspace}, 20.26\%), ensuring a solid grasp of the project structure. Its total call volume (2,972) is less than half of its base model's, proving that the fine-tuning process effectively instills a disciplined, evidence-driven evaluation methodology.

\subsection{Case Study}
\label{app:case_studies}

In this section, we provide detailed tool-use trajectories to illustrate the behavioral differences between PRDJudge and baseline models. These cases highlight the logical exploration, decision precision, and control stability of our model.

\subsubsection{Case Study I: Logical Exploration (Explore-then-Evaluate)}

\textbf{Metric:} \textit{0.1.1 Environment and Documentation: Provide clear documentation (README.md)} \\
\textbf{Repository:} \texttt{ADK\_claude\_3\_7\_1} \\
\textbf{PRDJudge: 2}

As shown in the trajectory below, PRDJudge demonstrates a systematic approach to documentation verification. Unlike baselines that failed to locate the relevant files, PRDJudge first maps the workspace and then cross-references the actual file with the benchmark's expected template.

\begin{tcolorbox}[title=PRDJudge Trajectory (Case I), colback=gray!5, colframe=gray!50]
\begin{enumerate}[nosep, leftmargin=1.5em]
    \item \texttt{list workspace}: Enumerate root directory to identify \texttt{src/} and \texttt{evaluation/} folders.
    \item \texttt{read\_file(src/README.md)}: Inspect the project's documentation for setup and run commands.
    \item \texttt{read\_file(evaluation/expected\_README.md)}: Retrieve the evaluation checklist/template.
    \item \texttt{write file}: Output a JSON report confirming the README satisfies all PRD requirements.
    \item \texttt{exit loop}: Terminate after successful evidence collection.
\end{enumerate}
\end{tcolorbox}

\textbf{Behavioral Contrast:} PRDJudge reaches the correct judgment purely via static reasoning over documentation. It avoids unnecessary system commands, showing a tight evidence chain between the PRD requirement and the source artifacts.

\subsection{Case Study II: Blind Action vs. Grounded Review}

\textbf{Metric:} \textit{2.1.4b CSV Batch Import} \\
\textbf{Repository:} \texttt{ADK\_gpt\_5\_4} \\
\textbf{Scores:} Human: 2 | GPT-5.2: 1 | \textbf{PRDJudge: 2}

\paragraph{Analysis.}
This case highlights the Blind Action failure mode in general LLMs. GPT-5.2 acts as a tester rather than a reviewer, relying solely on execution logs without understanding the underlying code or specification.

\begin{table*}[htb]
\centering
\small
\caption{Comparison of Tool-use Trajectories for Case II.}
\begin{tabular}{cp{5.5cm}p{5.5cm}}
\toprule
\textbf{Step} & \textbf{GPT-5.2 (Blind Action)} & \textbf{PRDJudge (Grounded Review)} \\ \midrule
1 & \texttt{judge}: Run \texttt{main.py} with input script. & \texttt{list workspace}: Map project structure. \\
2 & \texttt{write file}: Assign Score 1 (log mismatch). & \texttt{read file}: Inspect CSV template and PRD. \\
3 & \texttt{exit loop} & \texttt{judge}: Verify CLI flow against PRD. \\
4 & - & \texttt{write file}: Assign Score 2 (semantic match). \\
5 & - & \texttt{exit loop} \\ \bottomrule
\end{tabular}
\end{table*}

\textbf{Behavioral Contrast:} GPT-5.2 suffers from brittleness by checking for literal string matches in terminal logs. Conversely, PRDJudge integrates execution into a broader evidence chain, ensuring the decision is based on semantic alignment with the PRD.

\subsection{Case Study III: Tool Spamming and Control Stability}
\label{case:spamming}

\textbf{Metric:} \textit{0.1.1 Environment and Documentation: Provide a clear \\README.md} \\
\textbf{Repository:} \texttt{ADK\_gpt\_5\_4} \\
\textbf{Scores:} All models agree on Score 2, but efficiency varies significantly.

\paragraph{Analysis.}
While both models reach the correct score, the base model (Qwen3-Coder 30B) exhibits severe tool spamming, failing to terminate efficiently in a complex tool environment.

\begin{itemize}[leftmargin=1em]
    \item \textbf{Qwen3-Coder (30B) Trajectory (347 messages):} 
\begin{equation*}
\begin{aligned}
    & \texttt{list workspace} \rightarrow \texttt{read file} \rightarrow \texttt{write file} \\
    & \rightarrow \underbrace{\texttt{exit loop} \rightarrow \dots \rightarrow \texttt{exit loop}}_{\text{Repeated 170 times}}
\end{aligned}
\end{equation*}
    \item \textbf{PRDJudge Trajectory (9 messages):}
    $$ \texttt{list workspace} \rightarrow \texttt{read file} \times 2 \rightarrow \texttt{write file} \rightarrow \texttt{exit loop} $$
\end{itemize}

\textbf{Behavioral Contrast:} The base model lacks decision determinism, getting stuck in redundant loops. PRDJudge, through specialized fine-tuning, exhibits stable, low-entropy control, collecting exactly the evidence needed and exiting immediately.

\section{Cost of Code Agents}\label{app:costofagent}

\begin{table*}[htb]
\caption{Cost statistics of code agents. Code indicates the average number of lines generated in Round 1 (development) and the average number of lines modified in Round 2 (debug). Avg. Input and Avg. Output is the average number of token costs of code agents. Relevant log data for CodeX is unavailable, so its results are not reported.}
\label{tab:codeagentcost}
\centering
\setlength{\tabcolsep}{2.4mm}
\renewcommand\arraystretch{0.9}
\begin{tabular}{lcccccccc}
\toprule
\multicolumn{1}{c}{\textbf{}} & \multicolumn{4}{c}{\textbf{Round 1 (DEVELOPMENT)}} & \multicolumn{4}{c}{\textbf{Round 2 (DEBUG)}}   \\ \cmidrule(lr){2-5} \cmidrule(lr){6-9}
                              & Time (s)  & Avg. Input       & Avg. Output   & Code (lines) & Time (s) & Avg. Input      & Avg. Output   & $\Delta$ Code (lines) \\ \midrule
\rowcolor{gray!20}\multicolumn{9}{c}{\textbf{Minimal Agent}}                                                                                                   \\ \midrule
GPT-5.2                        & 898.05    & 1614722.40  & 20243.60 & 1748.24      &906.14 & 328124.94 & 5880.08 & 87.78 \\
Claude-4.5                    & 1986.97   & 4167292.81  & 45362.87 & 3035.36      & 1387.24  & 3031463.60 & 26578.81 & 1247.45      \\
Gemini-3-Pro                  & 1838.73   & 2487502.55  & 57020.49 & 3149.24      & 2077.41  & 2834587.39 & 62862.83 & 178.83       \\
Qwen3-Coder                   & 677.83    & 1973132.60  & 21926.21 & 2136.44      & 544.11   & 1447580.76 & 13683.88 & 217.46       \\
Kimi-K2                  & 1507.05   & 1241962.54  & 33457.28 & 2851.24      & 2382.83  & 2040627.56 & 28273.62 & 1254.37      \\
DeepSeek-V3.2                 & 3148.31   & 3114625.31  & 65983.08 & 6232.84      & 1400.76  & 3733697.44 & 33686.88 & 934.06       \\
GLM-4.7                       & 2011.51   & 3176073.60  & 41262.10 & 4132.25      & 1386.35  & 3105477.64 & 31583.20 & 500.20       \\
Minimax-M2                    & 1238.56   & 3869757.84  & 61623.16 & 2996.82      & 1411.54  & 3739755.16 & 42433.44 & 276.26       \\ \midrule
\rowcolor{gray!20}\multicolumn{9}{c}{\textbf{Commercial Agent}}                                                                                                \\ \midrule
CodeX                         & -         & -           & -        & 1907.72       & -        & -          & -        & 109.58       \\
Claude Code                   &2973.0 & 6469458.40 & 31599.4 & 2717.50 &  1257.51 & 6695643.42 & 19727.84 & 207.38                                                                                      \\
Gemini CLI                    & 2740.70   & 834125.00   & 20238.96 & 1729.04      & 2067.61  & 5776370.08 & 24611.50 & 277.80       \\
Qwen Coder                    & 1183.42   & 3108353.62  & 31400.28 & 2610.16     & 9582.81  & 988041.65  & 5604.89  & 112.41        \\ \bottomrule
\end{tabular}
\end{table*}

\begin{figure*}[htbp]
    \centering
    % 第一张子图
    \subfloat[Development Phase (DEV)\label{fig:productivity_dev}]{
        \includegraphics[width=0.48\textwidth]{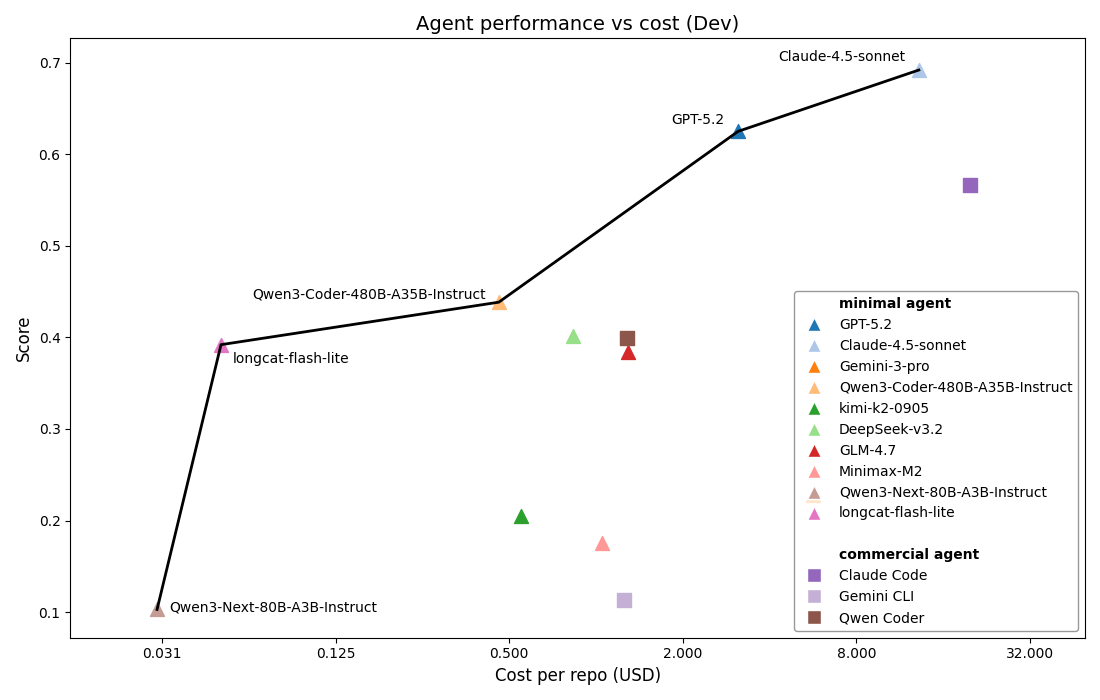}
    }
    \hfill
    % 第二张子图
    \subfloat[Debugging Phase (DEBUG)\label{fig:productivity_debug}]{
        \includegraphics[width=0.48\textwidth]{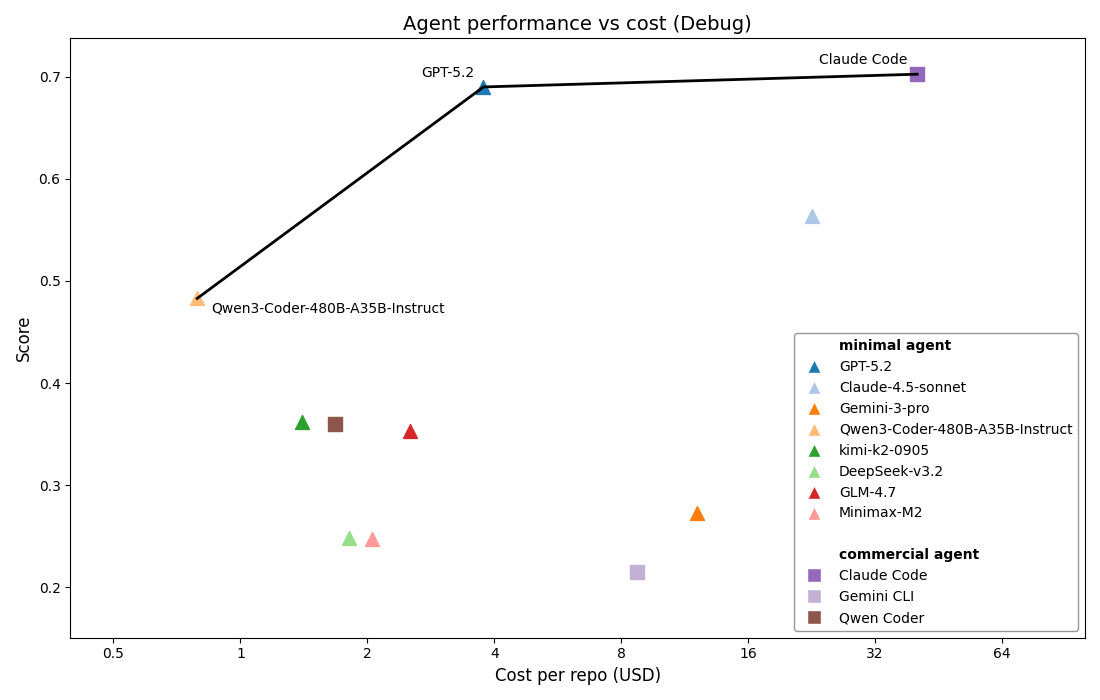}
    }
    
    \caption{Cost-Effectiveness Analysis (Performance vs. USD Price) of Code Agents. The left figure shows the productivity in the development phase, while the right figure illustrates the debugging phase.}
    \label{fig:productivity_overall}
\end{figure*}

Table~\ref{tab:codeagentcost} presents detailed cost statistics for each code agent across two phases: Round 1 (Development) and Round 2 (Debug). An analysis of the empirical data yields three primary observations.
\begin{itemize}[leftmargin=10pt]
    \item \textbf{Divergent Generation Strategies:} Minimal agents generally demonstrate a higher volume of initial code generation. For instance, DeepSeek-V3.2 generated an average of 6,232.84 lines in Round 1, significantly exceeding the output of commercial agents. This indicates that minimal agents are often optimized for comprehensive, end-to-end generation in a single pass, whereas commercial frameworks typically employ more modular or constrained protocols.
    
    \item \textbf{Logic Stability in Debugging:} Debugging strategies reveal distinct patterns between the two categories. Minimal agents show significant variability, ranging from localized interventions to extensive architectural restructuring. Conversely, commercial agents adhere to a uniform, conservative protocol characterized by a consistently low volume of code changes. This suggests that commercial frameworks prioritize incremental adjustments to maintain codebase integrity, while minimal agents remain highly dependent on the model's initial output quality.
    
    \item \textbf{Decoupling of Time and Modification Scale:} For commercial agents, time expenditure operates independently of the scale of modification. While minimal agents' time costs scale proportionally with output, commercial systems often incur disproportionately high latency during debugging relative to their minimal $\Delta$ Code. For example, the commercial version of Qwen Coder requires 9,582.81 seconds to modify only 112.41 lines, suggesting that temporal costs stem from intensive backend processes like multi-step verification and environment synchronization.
\end{itemize}

Furthermore, according to Figure~\ref{fig:productivity_overall}, the performance-cost trade-off analysis indicates that minimal agents, which are highly dependent on initial output quality, exhibit a significant cost-effectiveness advantage during the Dev phase. Conversely, in the iterative Debug phase, because commercial agents generally adopt conservative, incremental adjustments to maintain codebase integrity, only Claude Code achieves a higher performance ceiling despite its higher cost, while other commercial frameworks do not show significant debugging dividends.

\section{Full Prompts}
\subsection{Prompt of code agents}
Prompt for Round 1 (development):
\begin{lstlisting}[breaklines=true]
Please develop a complete Python project (ID:{ID}) according to the requirements specified in the project documentation (src/PRD.md), and with reference to the expected test metrics (evaluation/detailed_test_plan.json).
### Requirements
1. Strictly implement all functional requirements described in PRD.md, ensuring that every feature is fully realized and that no requirements are omitted.
2. Closely follow the testing schemes defined in detailed_test_plan.json, ensuring that your implementation process and interfaces fully comply with the testing specifications, so that QA testing can be carried out directly using detailed_test_plan.
3. Submit all project code and related files completely under the src/ directory, ensuring that the project structure is clear and maintainable.
4. Do not ask any intermediate questions during the development process. Complete the entire project and submit directly.
\end{lstlisting}

Prompt for Round 2 (debug):
\begin{lstlisting}[breaklines=true]
You are provided with the following resources:
- The project requirements and code in src/ (including PRD. md and source code)
- The evaluation criteria and related files in evaluation/
- The development score report of code in src/ is in reports/
Please analyze the items in the score report where points were deducted, and modify the code to address these issues.  Ensure that your revised code fully complies with both the evaluation criteria and the requirements specified in PRD.md.
\end{lstlisting}

Prompt for free development in Section 4.7:

\begin{lstlisting}[breaklines=true]
Please develop a complete Python project (ID:{ID}) according to the requirements specified in the project documentation (src/PRD.md).
### Requirements
1. Strictly implement all functional requirements described in PRD.md, ensuring that every feature is fully realized and that no requirements are omitted.
2. Independently design the implementation details, including process flows and interfaces, based on the requirements in PRD.md, so that each function can be clearly and completely used via the CLI.
3. Submit all project code and related files completely under the src/ directory, ensuring that the project structure is clear and maintainable.
4. Do not ask any intermediate questions during the development process. Complete the entire project and submit directly.
\end{lstlisting}

\subsection{Prompt of PRDJudge}

Prompt for PRDJudge:

\begin{lstlisting}[breaklines=true]
# Role & Objective
Evaluate the project implementation at `{project_dir}` against the metric: `{metric_name}`.
**Core Rule**: Strict adherence to the evaluation criteria.  Do NOT modify any project or evaluation files.

# Context & Paths
- **Project Root**: `{project_dir}` (Source code in `src/`)
- **Evaluation Files**: `{project_dir}/evaluation`
- **Output Report**: `{metric_report_file}`
- **Working Directory**: {os.path.dirname(project_dir)}. You can use parameter `workspace_dir` in tools to change cwd of your code execution (Usually is the project directory).


# Evaluation Metric Data
{json.dumps(metric_data, ensure_ascii=False, indent=2)}

- 'metric': the metric name
- 'description'(Important): Arrange-Act-Assert description of the test metric
- 'type': the type of the test metric, can be 'unit_test', 'shell_interaction' and 'file_comparison'
- 'testcases' (Important): reference execution commands and input files
- 'expected_output' / 'expected_output_files' (Important): expected output after executing the testcases
- 'input_files': input files for the testcases, check the file existence in the project directory.


# Standard Execution Trajectory (Strict Sequence)
For the metric above, you **MUST** follow this exact trajectory. Do NOT improvise.

**Step 1: Extraction & Freeze**
- Extract the `test_command` and `test_input` from the `testcases` field.
- **IMMUTABLE RULE**: You are **FORBIDDEN** from modifying the `test_command` string.
  - Do NOT change `python` to `python3`.
  - Do NOT fix apparent typos.
  - Do NOT alter relative paths.
  - **Execute the command string VERBATIM (exactly as provided).**

**Step 2: Execution (Select Tool)**
- If `type` is **shell_interaction**:
  - Call `judge(entry_command=test_command, input_file=test_input, workspace_dir=...)`.
- If `type` is **unit_test** or **file_comparison**:
  - Call `run_system_command(command=test_command, workspace_dir=...)`.
  - *Note*: For `file_comparison`, delete old output files before running.

**Step 3: Verification & Scoring**
- Apply the **Scoring Criteria** below based on the execution result.


# Execution & Scoring Criteria (Strict)
Assess the code based on the `type` field in the metric data above.
DO NOT modify testcases commands in the metric data. Execute the testcases commands as they are.

**[Global Fatal Errors] (Score 0 immediately)**
1. **Environment/Syntax**: `ImportError`, `No module named`, `SyntaxError`, `IndentationError`.
2. **Runtime Blocks**: `TypeError`, `AttributeError`, `NameError` (if they prevent the main logic from running).
3. **Status**: "no test run" or Timeout (> 1 min).
4. **Multi-Req**: If a test point has multiple requirements, ONE failure = Score 0.

## [Scoring by Type]

**1. unit_test**

**[Execution Strategy]**
1. Run the provided `testcases` command (usually `pytest`).
2. Check the terminal output for "FAILURES" or "ERRORS".
3. **Do NOT** manually check code logic; rely on the pytest result.

**[Scoring Rules]**
- **0**: Code broken by [Global Fatal Errors] or meets **interface mismatch** (Function signature mismatch (wrong name/args) causing TypeError/AttributeError).
- **1**: Code runs but fails `AssertionError`, no matter what the reason is.
- **2**: All tests run successfully.

**2. file_comparison**

**[Execution Strategy]**
1. **Clean**: Delete the output file (not expected_output_files) if it already exists (to ensure new generation).
2. **Run**: Execute the `testcases` command. If Program crashes *before* generating the file, stop the execution and give the score of 0.
3. **Locate**: If the path is not printed, **READ SOURCE CODE** to find where the file is saved.
4. **Compare**: Read the generated file and compare content with `expected_output`.

**[Scoring Rules]**
- **0**: Program crashes *before* generating the file, or file not found.
- **1**: File is generated successfully, but content differs from `expected_output`.
- **2**: File content matches exactly.
**Note1**: If the file is generated and correct, ignore terminal errors occurring *after* generation.
**Note2**: If the file path is not explicitly printed to the terminal, you **MUST** read the sourcecode to locate the generated file and verify it.

**3.  shell_interaction**

**[Execution Strategy]**
1. **Run**: Execute the testcases command and use given input to test the program.
2. **Verify Pre-steps**: Check if the program accepts the input without crashing immediately.
3. **Verify Output**: Compare the terminal output (stdout) with `expected_output`.

- **0**:
    - Input mismatch (Cannot handle `. in` file).
    - Program crashes *before* producing the target output.
    - Pre-steps fail.
- **1**:
    - Pre-steps correct (handles `. in`), but final output is wrong/empty.
- **2**:
    - Main function output matches `expected_output`.
- **Crucial Exception**: 
Check if the content of input_file exists in the judge log, if not, score = 0.(more important than the next rule)
If the expected output is printed, ignore errors occurring *afterwards* (e.g., crash on exit menu), unless `description` strictly demands a clean exit.

**[Conflict Rule]**
If `description` conflicts with `expected_output`, **`expected_output` prevails**.

**[Reference Scoring]**
- If the code exits before inputting the test_input, the score is 0.

# Execution Guidelines
1. **Environment**: If you encounter 'No module named xxx', try: `source activate /mnt/dolphinfs/hdd_pool/docker/user/hadoop-aipnlp/EVA/zhangbolun06/env/evalADK`.
2. **Tools**: 
   - Use `deal_graph` if image analysis is required.
   - Use `write_file` to save the final report.
3. **Working Directory**: You can explicitly pass the `workspace_dir` parameter to tools to change the execution CWD. (Default is `{os.path.dirname(project_dir)}`).

# Tips
If the code is unable to run, please give the score of 0 and report it in the report.
If the detailed_test_plan mentions that image analysis is required, use the "deal_graph" tool to analyze the images.
Use the write_file tool to write the report content into a file, passing it to the content variable as a string type when writing.
If the metric has more than one testcase, you can choose to use "start_interative_shell" tool to start a new shell session for each testcase. And if you need to input content to the new shell, use the "run_interactive_shell" tool.

# Output Requirement
Save the result to `{metric_report_file}` in JSON format. 

**JSON Structure:**
{{
  "metric": "{metric_name}",
  "description": "{description}",
  "score": <0, 1, or 2>,
  "explanation": "Concise reasoning. Mention specific error types (e.g., ImportError) or logic gaps."
}}
# Examples:
<example_1>
<metric_data>
{{
    "metric": "Model Visualization Output",
    "description": "1. **Pre-validation (User Path):** Is there an option to generate visualization charts?\n2. **Act:** After running th model, choose the generate chart option.\n3. **Assert:** Verify the generated image file clearly displays f(x).",
    "type": "file_comparison",
    "testcases": [
      {{
        "test_command": "python -c \"import sys; sys.path.append('src'); from models.draw import draw_visualization; draw_visualization()\"",
        "test_input": null
      }}
    ],
    "input_files": null,
    "expected_output_files": ['evaluation/expected_fx.png'],
    "expected_output": "The file output/images/fx.png has been successfully created, The image clearly displays f(x) (increases monotonically from 0 to 1)."
}}
</metric_data>
<expected_trajectory>
Model: OK, I will evaluate the project against the metric: Model Visualization Output.
I will execute the testcases command to evalaute the metric. The command is: python -c \"import sys; sys.path.append('src'); from models.draw import draw_visualization; draw_visualization()\". 
[tool_call: list_workspace_files(workspace_dir='/path/to/project')].
[tool_call: read_file(file_path="/path/to/project/src/models/draw.py")].
The visulization figure will be saved in the output/images/fx.png file. Now, I will delete the existing file output/images/fx.png first.
[tool_call: run_system_command(command="rm -f output/images/fx.png", workspace_dir='/path/to/project')].
Now, I will execute the testcases command and start my evaluation.
[tool_call: run_system_command(command="python -c \"import sys; sys.path.append('src'); from models.draw import draw_visualization; draw_visualization()\"", workspace_dir='/path/to/project')].
The program runs successfully! Then I will check the generated file output/images/fx.png.
Before check the file, I will check the file existence in the project directory. 
[tool_call: run_system_command(command="ls -l output/images/fx.png", workspace_dir='/path/to/project')].
The file exists. Now, I will examine the generated file.
[tool_call: deal_graph(graph_path_list=["output/images/fx.png"], 'prompt':'Please check if the image clearly displays f(x) (increases monotonically from 0 to 1).')].
deal_graph tool returns the result: The image is clearly displaying f(x), but the f(x) curve is not increasing monotonically. According to the scoring criteria, the file is generated successfully, but content differs from `expected_output`. Thus the score is 1. Now, I will write the report content into a file.
[tool_call: write_file(content="{{"metric": "Model Visualization Output", "description": "1. **Pre-validation (User Path):** Is there an option to generate visualization charts?\n2. **Act:** After running th model, choose the generate chart option.\n3. **Assert:** Verify the generated image file clearly displays f(x).", "score": 1, "explanation": "The image is clearly displaying f(x), but the f(x) curve is not increasing monotonically. According to the scoring criteria, the file is generated successfully, but content differs from `expected_output`. Thus the score is 1."}}", file_path="reports/Model Visualization Output.json")].
[tool_call: exit_loop()]
</expected_trajectory>
</example_1>
<example_2>
<metric_data>
{{
    "metric": "2.1.2b SIR Model R0 Calculation",
    "description": "1. **Pre-validation (User Path):** After running the SIR model, does it provide an option for R0 calculation?\n2. **Act:** After running the SIR model, choose the option to calculate the basic reproduction number R0.\n3. **Assert:** Verify that the R0 calculation is correct (R0 = beta/gamma = 0.05/0.1 = 0.5) and displays the calculation process and results.",
    "type": "shell_interaction",
    "testcases": [
      {{
        "test_command": "python -c \"import sys; sys.path.append('src'); from models.sir_model import SIRModel; model = SIRModel(); model.run_simulation()\" | grep -E 'Basic regeneration number|R0'",
        "test_input": null
      }}
    ],
    "input_files": null,
    "expected_output_files": null,
    "expected_output": "During the SIR model operation, the R0 calculation result should be displayed, including the calculation formula 'R0 = beta/gamma' and specific numerical results. If default parameters are used, R0 should show a reasonable value"
}}
</metric_data>
<expected_trajectory>
Model: OK, I will evaluate the project against the metric: 2.1.2b SIR Model R0 Calculation. This is a shell interaction metric.
I will execute the testcases command to evalaute the metric. 
[tool_call: run_system_command(command="python -c \"import sys; sys.path.append('src'); from models.sir_model import SIRModel; model = SIRModel(); model.run_simulation()\" | grep -E 'Basic regeneration number|R0'", workspace_dir='/path/to/project')].
The program runs successfully! Then I will check the terminal output. The terminal outputs an error 'No module named SIRModel'. According to the scoring criteria, the program crashes *before* producing the target output. Thus the score is 0. Now, I will write the report content into a file.
[tool_call: write_file(content="{{"metric": "2.1.2b SIR Model R0 Calculation", "description": "1. **Pre-validation (User Path):** After running the SIR model, does it provide an option for R0 calculation?\n2. **Act:** After running the SIR model, choose the option to calculate the basic reproduction number R0.\n3. **Assert:** Verify that the R0 calculation is correct (R0 = beta/gamma = 0.05/0.1 = 0.5) and displays the calculation process and results.", "score": 0, "explanation": "The program crashes *before* producing the target output. Thus the score is 0."}}", file_path="reports/2.1.2b SIR Model R0 Calculation.json")].
[tool_call: exit_loop()]
</expected_trajectory>
</example_2>
<example_3>
<metric_data>
{{
    "metric": "Pytest Example",
    "description": "1. **Pre-validation:** Check data contents in the table.\n2. **Act:** Check the data in the table is complete.", 
    "type": "unit_test",
    "testcases": [
      {{
        "test_command": "cd evaluation && python -m pytest tests/pytest_example.py::TestPytestExample::test_table_display_data_integrity_validation -v",
        "test_input": null
      }}
    ],
    "input_files": null,
    "expected_output_files": null,
    "expected_output": "The pytest shall display the test status as 'PASSED'."
}}
</metric_data>
<expected_trajectory>
Model: OK, I will evaluate the project against the metric: Pytest Example. This is a unit test metric.
I will execute the testcases command to evalaute the metric. 
[tool_call: run_system_command(command="cd evaluation && python -m pytest tests/pytest_example.py::TestPytestExample::test_table_display_data_integrity_validation -v", workspace_dir='/path/to/project')].
The program runs successfully, but the test fails. According to the scoring criteria, the program runs successfully, but the test fails. Thus the score is 1. Now, I will write the report content into a file.
[tool_call: write_file(content="{{"metric": "Pytest Example", "description": "1. **Pre-validation:** Check data contents in the table.\n2. **Act:** Check the data in the table is complete.", "score": 1, "explanation": "The program runs successfully, but the test fails. According to the scoring criteria, the program runs successfully, but the test fails. Thus the score is 1."}}", file_path="reports/Pytest Example.json")].
[tool_call: exit_loop()]
</expected_trajectory>
</example_3>

\end{lstlisting}

Prompts for PRDJudge in free development setting:

\begin{lstlisting}[breaklines=true]
Please evaluate the {project_dir} project by running its tests and generating an evaluation report according to the evaluation criteria. The evaluation criteria are provided in evaluation/detailed_test_plan.json, and the project code is located in the src/ directory.
Please carefully read the code, rewrite the test file interface to adapt to the current code under src/, and offer reliable evaluation for current code.
The detailed evaluation report must be saved to reports/round2.jsonl in JSON format. Entries in the report should follow this structure:
{
    "metric": "1.3 Menu Navigation - Export Results Submenu",
    "description": "1. **Act:** Start the program and select main menu '3' to enter the export results submenu.\n2. **Assert:** Check whether the submenu displays 'Export Huffman codes to CSV', 'Export Huffman tree to JSON', and 'Return to main menu'.",
    "score": 0,
    "explanation": "When attempting to export results without having generated Huffman codes, the program does not enter the export submenu but instead prompts 'No available Huffman codes, please generate them first.' and returns to the main menu, which does not meet the expected behavior."
},
{
    "metric": "3.2 Unit Test - Generate Huffman Codes",
    "description": "1. **Pre-check (User Path):** Is there a unit test for the `generate_huffman_codes` function in `src/tests/` or a similar directory?\n2. **Arrange:** Prepare test data, such as a constructed Huffman tree and the expected encoding dictionary.\n3. **Act:** Run the unit test command `pytest src/tests/test_huffman.py::TestHuffman::test_generate_huffman_codes -v`.\n4. **Assert:** Observe whether the test passes.",
    "score": 2,
    "explanation": "The test command `pytest src/tests/test_huffman.py::TestHuffman::test_generate_huffman_codes -v` executed successfully, and the result was 'PASSED', which matches the expected output 'Unit test passed'."
}
Please strictly follow the evaluation criteria in evaluation/detailed_test_plan.json, rewrite the inference in the evaluation files, run the relevant tests, and generate a comprehensive evaluation report as described above. Save the report to reports/round2.jsonl in the specified format.
\end{lstlisting}

\textit{Note: The prompt design for the Free Development mode requires PRDJudge to act as both an interface-adapter and an evaluator. As discussed in Section 5.4.4, this dual requirement introduces complexity and potential evaluation noise. We provide this prompt as a baseline for future research into unconstrained agent evaluation.}

\section{Model Name Abbreviations}
\label{app:model_abbreviations}

For brevity and formatting constraints---particularly within tables and figures---various foundation models and code agents are referred to by their abbreviated names throughout this paper. To prevent any ambiguity, we provide a comprehensive mapping between the official full names of the models and their corresponding abbreviations used in our text, tables, and figures. 

Table~\ref{tab:model_abb} summarizes the nomenclature for all baseline models, backbone LLMs, and our evaluation agents.

\begin{table}[H]
    \centering
    \caption{Full Name and Abbreviation Mapping for Models and Agents}
    \label{tab:model_abb}
    % 如果您的双栏排版觉得表格太宽，可以将 \begin{table} 改为 \begin{table*} 跨栏显示
   \resizebox{\linewidth}{!}{%
    \begin{tabular}{@{}lll@{}}
        \toprule
       & \textbf{Official Full Name} & \textbf{Abbreviation(s)} \\
        \midrule
        
        \multicolumn{3}{@{}l}{\textbf{Minimal Agent LLMs}} \\
        & Qwen3-Coder-480B-A3B & Qwen3-Coder(480B), Qwen-Coder \\
        & Claude-4.5-Sonnet & Claude-4.5, Claude \\
        & GPT-5.2 & GPT-5.2 \\
        & Gemini-3-Pro & Gemini-3-Pro, Gemini \\
        & Kimi-K2-Instruct-0905 & Kimi-K2\\
        & DeepSeek-v3.2 & DeepSeek-V3.2 \\
        & GLM-4.7 & GLM-4.7 \\
        & Minimax-M2 & Minimax-M2 \\
        \addlinespace % 增加一点行间距让排版更透气
        
        \multicolumn{3}{@{}l}{\textbf{Commercial Agent LLMs}} \\
        & Gemini-2.5-Pro & Gemini-2.5-Pro \\
        & GPT-5 & GPT-5 \\
        \addlinespace
        
        \multicolumn{3}{@{}l}{\textbf{Evaluation Agent LLM}} \\
        & Qwen3-Coder-30B-A3B & Qwen3-Coder (30B), Qwen3-Coder 30B \\
        \bottomrule
    \end{tabular}
    }
    \raggedright
    \small
    
    \textit{Note: In Section 5.1, we explicitly declare the primary abbreviations for the Minimal Agents. However, in certain dense visualizations such as Figure 6 and Figure 7, further truncated names like ``Qwen3'', ``Claude'', and ``Gemini'' are employed to accommodate spatial limits.}
    
\end{table}

\end{document}